\documentclass[preprint,12pt]{elsarticle}

\usepackage{amssymb}

\usepackage{hyperref}
\usepackage{bm}
\usepackage{amsmath}

\usepackage{listings}
\usepackage{xcolor}
\usepackage{tabularx}

\definecolor{codegreen}{rgb}{0,0.6,0}
\definecolor{codegray}{rgb}{0.5,0.5,0.5}
\definecolor{codepurple}{rgb}{0.58,0,0.82}
\definecolor{backcolour}{rgb}{0.95,0.95,0.92}

\lstdefinestyle{mystyle}{
    backgroundcolor=\color{backcolour},   
    commentstyle=\color{codegreen},
    keywordstyle=\color{magenta},
    numberstyle=\tiny\color{codegray},
    stringstyle=\color{codepurple},
    basicstyle=\ttfamily\footnotesize,
    breakatwhitespace=false,         
    breaklines=true,                 
    captionpos=b,                    
    keepspaces=true,                 
    numbers=left,                    
    numbersep=5pt,                  
    showspaces=false,                
    showstringspaces=false,
    showtabs=false,                  
    tabsize=2
}

\newcounter{bla}

\journal{Elsevier}

\begin{document}

\begin{frontmatter}



\title{JAX-SSO: Differentiable Finite Element Analysis Solver for Structural Optimization and Seamless Integration with Neural Networks}


\author[a]{Gaoyuan Wu\corref{author}}

\cortext[author] {Corresponding author.\\\textit{E-mail address:} gaoyuanw@princeton.edu}
\address[a]{Department of Civil and Environmental Engineering, Princeton University, Princeton, NJ 08544, United States}

\begin{abstract}

Differentiable numerical simulations of physical systems have gained rising attention in the past few years with the development of automatic differentiation tools. This paper presents JAX-SSO, a differentiable finite element analysis solver built with JAX, Google’s high-performance computing library, to assist efficient structural design in the built environment. With the adjoint method and automatic differentiation feature, JAX-SSO can efficiently evaluate gradients of physical quantities in an automatic way, enabling accurate sensitivity calculation in structural optimization problems. Written in Python and JAX, JAX-SSO is naturally within the machine learning ecosystem so it can be seamlessly integrated with neural networks to train machine learning models with inclusion of physics. Moreover, JAX-SSO supports GPU acceleration to further boost finite element analysis. Several examples are presented to showcase the capabilities and efficiency of JAX-SSO: i) shape optimization of grid-shells and continuous shells; ii) size (thickness) optimization of continuous shells; iii) simultaneous shape and topology optimization of continuous shells; and iv) training of physics-informed neural networks for structural optimization. We believe that JAX-SSO can facilitate research related to differentiable physics and machine learning to further address problems in structural and architectural design.

\end{abstract}


\end{frontmatter}



\section{Introduction}\label{sec:intro}
Structural optimization is of great significance in finding efficient shapes, sizes of structural elements, and topology of structures that lead to efficient material usage, providing the built environment with sustainable structural and architectural design. Gradient-based methods have been implemented widely for structural optimization purposes \cite{Adriaenssens2014Shell,Bletzinger2001Structural}. Traditionally, analytical derivatives or numerical differentiation has to be derived in order to leverage gradient-based optimization methods \cite{Keulen2005Review}. However, such processes can either be tedious, error-prone or less accurate, making it difficult to develop an automated framework to assist structural optimization. \par
Differentiable numerical simulations of physical systems have drawn increasing attention in the past few years. Differentiable simulations make use of automatic differentiation tools so that the derivatives of physical quantities with respect to input variables can be calculated in an automated manner, saving the burden of analytical derivation and avoiding the inaccuracy brought by numerical differentiation \cite{Baydin2018Automatic}. Thus, differentiable simulations can not only be implemented to solve forward problems, but can also be used to solve inverse problems, such as optimization problems \cite{Bezgin2023JAX-Fluids:,Chandrasekhar2021AuTO:,Chandrasekhar2021TOuNN:,Chandrasekhar2023FRC-TOuNN:,Pastrana2023JAX,Wu2023framework,Schoenholz2021JAX,Xue2023JAX-FEM:}. Moreover, differentiable simulations can be effortlessly integrated with machine learning algorithms, such as artificial neural networks, enabling exciting research at the intersection of machine learning and numerical simulations \cite{Bezgin2023JAX-Fluids:,Kochkov2021Machine,Schoenholz2021JAX}. \par
In the field of structural engineering and architectural design, differentiable solvers have been emerging: Pastrana et al. \cite{Pastrana2023JAX} developed a differentiable solver for form finding based on the force density method (FDM); Wu \cite{Wu2023framework} proposed a framework for structural shape optimization based on automatic differentiation; Chandrasekhar et al. \cite{Chandrasekhar2021TOuNN:} proposed a differentiable framework for topology optimization; Hoyer et al. \cite{Hoyer2019Neural} proposed a structural optimization technique incorporating neural networks with the help of differentiable simulations. Xue et al. \cite{Xue2023JAX-FEM:} developed JAX-FEM, a differentiable finite element solver, but it is intended for computational mechanics of mechanical structures instead of architectural design. Our previous research on differentiable finite element solver \cite{Wu2023framework} showcases how differentiable physics can be implemented to solve shape optimization problems of grid-shells, but there are a few limitations: i) it does not support arbitrary objective functions: it can only be used for strain energy minimization; ii) it only supports shape optimization problems but not size or topology optimization problems; iii) it only supports beam elements; and iv) it does not showcase how differentiable solvers can be integrated with machine learning in the context of structural optimization and design. The research gaps are thus identified: the field of structural and architectural engineering lacks a differentiable finite element solver to assist generic gradient-based optimization and physics-informed machine learning research based on differentiable simulations of structures. \par
In response, this paper presents JAX-SSO, a differentiable solver for finite element analysis (FEA), to enable better design for the built environment. The solver is based on JAX \cite{2023JAX:}, Google’s high-performance computing library for machine learning research. JAX has been successfully implemented for research in differentiable computational fluid dynamics \cite{Bezgin2023JAX-Fluids:,Kochkov2021Machine}, molecular dynamics \cite{Schoenholz2021JAX}, computational mechanics \cite{Xue2023JAX-FEM:}, and structural optimization \cite{Pastrana2023JAX,Wu2023framework}. We highlight the following features of JAX-SSO: 
\begin{enumerate}
    \item Automatic derivatives evaluations using automatic differentiation (AD) and the adjoint method \cite{Keulen2005Review}. This feature comes in handy when conducting structural optimization problems. The gradient of arbitrary objective function with respect to design variables can be evaluated effortlessly. 
    \item Seamless integration with machine learning libraries. Written in Python and JAX, JAX-SSO is naturally within the machine learning ecosystem. With the AD feature of JAX-SSO, physics-informed neural networks can be trained easily since the gradients needed in backpropagation are easy to obtain.
    \item Vectorized code structure and GPU acceleration for faster simulations and derivatives evaluations.
    \item Various element types: truss, beams and shells based on MITC-4 formulation \cite{Dvorkin1984continuum}. 
\end{enumerate}
The remaining of this paper is outlined as follows. Section \ref{sec:jaxsso} introduces the basics of JAX-SSO: how JAX-SSO solves linear finite element analysis problems; how it can be used to solve optimization problems; and how it can be integrated with neural networks to assist structural optimization. The validation and performance assessment of JAX-SSO are also included in Section \ref{sec:jaxsso}. Section \ref{sec:ex} presents several examples to showcase the capabilities of JAX-SSO for structural optimization and integration with neural networks. This paper is concluded in Section \ref{sec:con}.

\section{Differentiable Finite Element Solver JAX-SSO}\label{sec:jaxsso}

\subsection{Features of JAX-SSO}
We first highlight some important methods and concepts used by JAX-SSO. 
\subsubsection{Automatic Differentiation}
For automatic derivatives evaluations, automatic differentiation (AD) \cite{Baydin2018Automatic} is implemented with the help of JAX. AD differentiates from analytical derivatives because it only outputs the numeric values of the derivatives instead of an analytical mathematical expression. AD also distinguishes from numerical differentiation because AD outputs analytical derivative values to the working precision of the computers instead of an approximation. For a complete introduction to AD, please refer to \cite{Baydin2018Automatic} . 
The variables in JAX-SSO are traceable: the program is aware of its full lifecycle within finite element analysis. Let us consider the construction of the global stiffness matrix $\bm{K}$. For each structural element in the structural system, the element’s attributes $\bm{x}$ (for instance, coordinates of the nodes, Young’s modulus, cross-section dimension, and Poisson’s ratio, etc.) determine the element’s stiffness matrix $\bm{k_e(x)}$, which further influences the global stiffness matrix $\bm{K}$. Because all the variables are \textit{traceable}, the partial derivative of the function output with respect to inputs can be obtained at every intermediate step, and through the \textit{chain rule}, the derivatives of $\bm{K}$ with respect to $\bm{x}$ can be automatically evaluated by AD: $\frac{\partial \bm{K}}{\partial \bm{x}} = \frac{\partial \bm{K}}{\partial \bm{k_e}}\frac{\partial \bm{k_e}}{\partial \bm{x}}$. The automatic differentiability of JAX-SSO facilitates gradient-based structural optimization and integration with the training of neural networks.

\subsubsection{Vectorized Code and Just-in-time Compilation}\label{sec:vecjit}

To fully exploit the capabilities of JAX, the solver is written in a vectorized manner instead of explicit for-loops. For instance, the assembly of the global stiffness matrix K from element’s stiffness matrix $\bm{k_e}$ is traditionally done in a for-loop, i.e., the contribution of each element to the global stiffness matrix is added to $\bm{K}$ in a sequential order. Here in JAX, we employ the concept of array programming \cite{Walt2011NumPy} and vectorize such operation so that the assembly of $\bm{K}$  is achieved simultaneously from each element. In addition to vectorization, JAX-SSO uses just-in-time (JIT) compilation. The speed of Python codes is often limited by the fact that high-level Python functions need to be interpreted into low-level machine codes and then executed. Through JIT compilation, key functions in JAX-SSO are compiled into machine codes at their first run and future calls of these functions will be much faster because no translation from high-level Python codes to low-level machine codes is needed.

\subsection{Finite element analysis with JAX-SSO}
\begin{figure*}[h]
\centering
\includegraphics[width=0.85\linewidth]{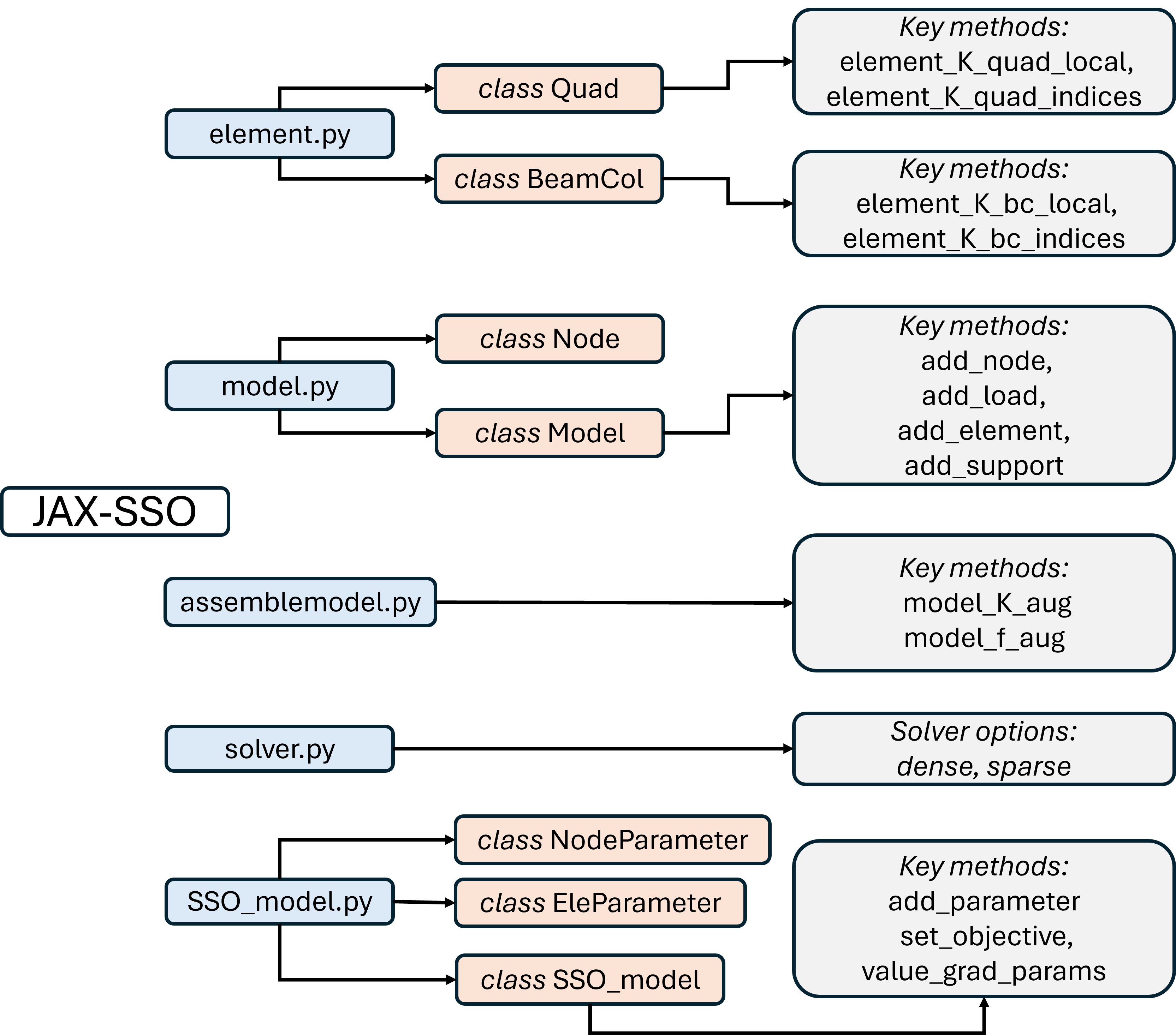}
\caption{The architecture of JAX-SSO package}
\label{fig:jaxsso}
\end{figure*}

We introduce the finite element analysis (FEA) procedure in JAX-SSO. In this paper we only consider solving linear static problems. Figure \ref{fig:jaxsso} illustrates the architecture of JAX-SSO package and some key methods that will be introduced in the following sections.

\subsubsection{Constructing the linear system of equations}
We first introduce how to construct the linear system of equations for FEA:
\begin{equation}\label{eq:ku=f}
        \bm{K}\bm{u}=\bm{f}
\end{equation}
Where $\bm{K}\in \mathbb{R}^{dof\times dof}$ is the global stiffness matrix where the superscript $dof$ denotes the number of degrees of freedom in the system, $\bm{f}\in \mathbb{R}^{dof}$ is the generalized nodal loading and $\bm{u}\in \mathbb{R}^{dof}$ is the nodal displacement vector which is the solution. 
Under \texttt{element.py} module (Figure \ref{fig:jaxsso}), different classes are defined for different element types: class beamcol for beam-columns and class quad for quadrilateral (quad) shell elements based on Mixed Interpolation of Tensorial Components-4 (MITC-4) formulation \cite{Bathe2006Finite}. Some key methods to these classes are functions that output local stiffness matrices: \texttt{element\_K\_bc\_local} for beam-column elements and \texttt{element\_K\_quad\_local} for quad shell elements. For instance, the method \texttt{element\_K\_quad\_local} returns local stiffness matrix $\bm{k}_{quad}\in \mathbb{R}^{24\times 24}$ based on the attributes of the quad element $\bm{a}_{quad} =
\begin{pmatrix}
\bm{x}_{quad} & \bm{y}_{quad} & \bm{z}_{quad} & t & E & \nu & \kappa_x &\kappa_y
\end{pmatrix}\in \mathbb{R}^{17}$ where $
\begin{pmatrix}
    \bm{x}_{quad} & \bm{y}_{quad} & \bm{z}_{quad}
\end{pmatrix}\in \mathbb{R}^{12}$ 
are the coordinates of four associated nodes, $t$ is the thickness, $E$ is Young’s modulus, $\nu$ is Poisson’s ratio, and $\kappa\_x$, $\kappa\_y$ are the stiffness modification coefficients for local X and Y axis, respectively. Another set of key methods to these classes, \texttt{element\_K\_bc\_indices} and \texttt{element\_K\_bc\_indices}, are functions that return corresponding row and column indices in the global stiffness matrix at which the values of local stiffness matrix of this element are assigned to.\par
The module model.py (Figure \ref{fig:jaxsso}) is used to construct the model for FEA. Nodes, elements, loads, and boundary conditions can be added using some key methods in \texttt{model.py}: \texttt{add\_node}, \texttt{add\_load}, \texttt{add\_element}, and \texttt{add\_support}. After all the elements are properly defined, one can leverage methods in \texttt{assemblemodel.py} to obtain the global stiffness matrix $\bm{K}$. Obtaining $\bm{K}$ in JAX-SSO takes advantage of vectorization as mentioned in Section \ref{sec:vecjit}: instead of going through every structural element in the model and add its local stiffness matrix to $\bm{K}$ one by one, we use JAX’s \texttt{vmap} to map the key methods in \texttt{element.py} into higher dimension functions so that they can operate on all structural elements simultaneously in a vectorized way. In JAX-SSO, to optimize the storage and computational time, the global stiffness matrix $\bm{K}$ is stored as a sparse matrix in Batched Coordinate (BCOO) format. After obtaining $\bm{K}$, we use Langrage Multiplier method to impose boundary conditions and the linear system of equations to solve read:
\begin{equation}\label{eq:kaug}
        \bm{K}_{aug}\bm{u}_{aug}=\bm{f}_{aug}
\end{equation}
where $\bm{K}_{aug}=\begin{bmatrix}
    \bm{K} & \bm{V}^T \\
    \bm{V} & \bm{0}
\end{bmatrix}\in \mathbb{R}^{(dof+dof_{bc})\times(dof+dof_{bc})}$ is the augmented global stiffness matrix; $\bm{V}\in \mathbb{R}^{dof_{bc} \times dof}$ is a matrix that imposes the boundary conditions on the displacement vector, $dof_{bc}$ is the number of degrees of freedom constrained by the boundary conditions; $\bm{f}_{aug}=\begin{pmatrix}
    \bm{f} \\
    \bm{b}
\end{pmatrix} \in \mathbb{R}^{(dof+dof_{bc})}$ is the augmented loading vector where $\bm{b}\in \mathbb{R}^{dof_{bc}}$ is a vector of imposed boundary conditions; $\bm{u}_{aug}\in \mathbb{R}^{dof + dof_{bc}}$ is the augmented displacement vector whose first dof-rows yield  $\bm{u}$. The terms related to the boundary conditions yield the following condition:
\begin{equation}\label{eq:bc_eq}
        \bm{Vu}=\bm{b}
\end{equation}
The key methods in to obtain $\bm{K}_{aug}$ and $\bm{f}_{aug}$ are \texttt{model\_K\_aug} and \texttt{model\_f\_aug} . Code snippet \ref{code:addmodel} shows how to build a finite element model in JAX-SSO.
\begin{lstlisting}[language=Python, label={code:addmodel}, caption={Create a model in JAX-SSO and conduct FEA}]
import JaxSSO.model as Model 
model = Model.Model() #Create a model object

for i in range(n_node):
    model.add_node(i,x_nodes[i],y_nodes[i],z_nodes[i]) # Add nodes
    if i not in design_nodes:
        model.add_support(i,[1,1,1,1,0,1]) #Add boundary conditions
    else:
        model.add_nodal_load(i,nodal_load=[0.0,0.0,-Q,0.0,0.0,0.0]) #Add loads
  
for i in range(n_ele):
    i_node = cnct[i,0]
    j_node = cnct[i,1]
    model.add_beamcol(i,i_node,j_node,E,G,Iy,Iz,J,A) #Add a beam column
 
model.model_ready()  #calls key methods of assemble_model to form linear system of equations to solve
model.solve(which_solver='sparse')  #solve FEA
\end{lstlisting}

\subsubsection{Solving the linear system of equations}
Table \ref{table:solvers} presents the solvers in JAX-SSO to obtain the solution to Equation \ref{eq:kaug}, which includes both dense solvers and sparse direct solvers. For the dense solver, we implement JAX’s \texttt{jax.numpy.linalg.solve} which uses lower-upper (LU) decomposition to solve for $\bm{u}_{aug}$ and it supports both CPU and GPU devices. For the sparse direct solvers, two options are provided: \texttt{jax.experimental.sparse.linalg.spsolve} which implements QR factorization and can only operate on GPU; \texttt{scipy.sparse.linalg} that involves sparse LU decomposition using UMFPACK algorithm \cite{Davis2004Algorithm} and can only operate on CPU. 
\begin{table*}[h]
\centering 
\caption{Solver options in JAX-SSO}\label{table:solvers}

\begin{tabular}{{l}{l}{l}}
Solver Options              & Sparse or Dense & Devices \\
\hline
\texttt{jax.numpy.linalg.solve} & Dense & CPU\&GPU \\ 
\texttt{jax.experimental.sparse} & Sparse & GPU \\
\texttt{.linalg.spsolve} & & \\
\texttt{scipy.sparse.linalg} & Sparse & CPU \\
\hline
\end{tabular}
\end{table*}
It should be noted that even though the supported devices for solving of the linear system of equations $\bm{K}_{aug} \bm{u}_{aug}=\bm{f}_{aug}$ depend on the selected solver, the assembly of the linear system of equations $\bm{K}_{aug} \bm{u}_{aug}=\bm{f}_{aug}$ can be done by either GPU or CPU, no matter which solver will be used to solve for the solution. These solver options are stored in module \texttt{solver.py} (Figure \ref{fig:jaxsso}).

\subsubsection{Validation of JAX-SSO for FEA}
\begin{figure*}[h]
\centering
\includegraphics[width=1.0\linewidth]{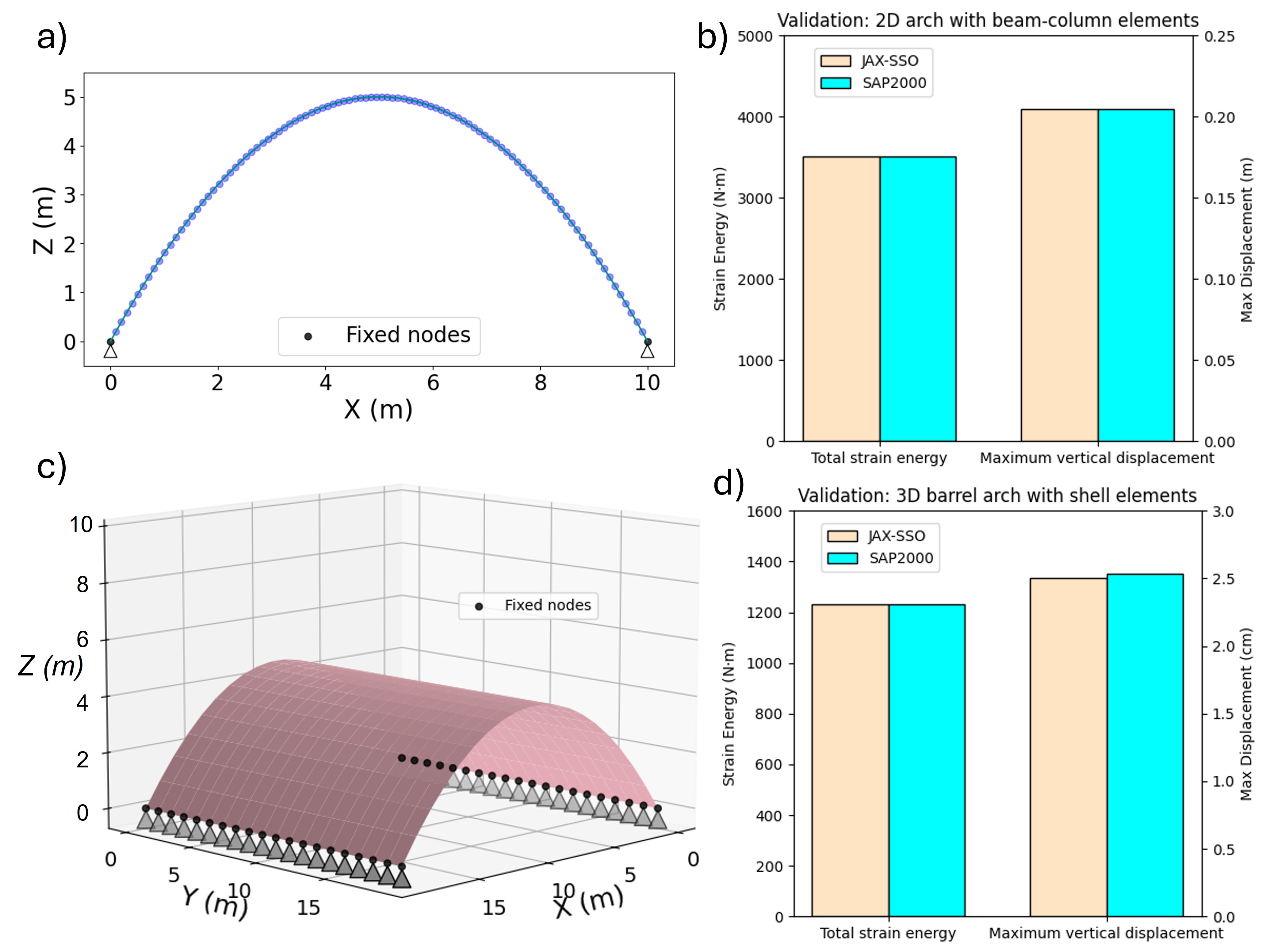}
\caption{Validation of JAX-SSO for FEA: a) 2D arch with beam-column elements; b) Comparison between JAX-SSO and SAP2000: 2D arch; c) 3D barrel arch with quad shell elements; d) Comparison between JAX-SSO and SAP 2000: 3D barrel arch}
\label{fig:fea_valid}
\end{figure*}

In this section, we validate our solver against commercial software SAP2000. The first example for validation is a simply supported 2D arch bridge that spans 10m, illustrated in Figure \ref{fig:fea_valid}.a. The arch bridge is simply supported on both ends and it has a parabola shape with a rise of 5m. The arch is discretized into 99 beam elements and downward nodal load with 500N magnitude is applied to each node. All the elements have the same properties: Young’s modulus $E=1.99\times10^8$Pa, Poisson’s ratio $\nu=0.3$, moment of inertia $I_y=6.6\times10^{-5}$ m$^4$ and $I_z=3.3\times10^{-6}$ m$^4$, and cross-sectional area $A=4.3\times10^{-3}$ m$^2$. Figure \ref{fig:fea_valid}.b presents the maximum nodal displacement and total strain energy (calculated as $0.5\bm{f}^T \bm{u}$) of the structure from JAX-SSO and SAP2000, which validates the solution from JAX-SSO. The percentage difference is 0.05\% for the maximum nodal vertical displacement along Z-direction and 0.06\% for the total strain energy. \par
The second validation example is a barrel arch consisting of quad shell elements, as can be seen in Figure \ref{fig:fea_valid}.c. The barrel arch has a parabolic shape and spans 19m on both X and Y axis. The height of the structure is 4.5m. Regular grids are used to discretize the structure into 400 quad shell elements and downward nodal load of 500 KN is applied to each node. All the shell elements have the same property: Young’s modulus $E=1.99\times10^8$ Pa, Poisson’s ratio $\nu=0.2$, and thickness $t = 0.25$m. The nodes on the ground are pin supported. Figure \ref{fig:fea_valid}.d illustrates the difference between JAX-SSO and SAP2000 in terms of the strain energy and maximum displacement: the total strain energy is 1229.7 N$\cdot$m from JAX-SSO and 1231.2 N$\cdot$m from SAP2000; the maximum vertical displacement along Z direction is 2.50 cm from JAX-SSO and 2.53 cm from SAP2000 with a difference of 1.33$\%$. The two examples above prove the validity of JAX-SSO.

\subsubsection{Performance of JAX-SSO for FEA}\label{sec:fea_per}
\begin{table}[h]
\centering 
\caption{FEA Options}\label{table:fea_options}

\begin{tabular}{{l}{l}{l}}
FEA options              & Assembly Device & Solver Option and Device \\
\hline
Scipy-sparse(CPU\&GPU) & GPU & \texttt{scipy.sparse.linalg}\\ 
& &  \texttt{.spsolve} on CPU\\
Scipy-sparse(CPU) & CPU & \texttt{scipy.sparse.linalg}\\\
& &  \texttt{.spsolve} on CPU\\
JAX-sparse(GPU) & GPU & \texttt{jax.experimental.sparse} \\ 
& & \texttt{.linalg.spsolve } on GPU\\
Dense(GPU) & GPU & \texttt{jax.numpy.linalg.solve}\\
& &  on GPU\\
Dense(CPU) & CPU & \texttt{jax.numpy.linalg.solve}\\
& &  on CPU\\
\hline
\end{tabular}
\end{table}
In this subsection, we present the performance of JAX-SSO in terms of its speed for conducting FEA. Different options for the assembly and solving of the linear system of equations $\bm{K}_{aug}\bm{u}_{aug}=\bm{f}_{aug}$ are considered. Table \ref{table:fea_options} presents the different FEA options of JAX-SSO. The performance tests are conducted on Princeton’s DELLA cluster with Nvidia’s A100 GPU (80 Gigabytes graphics memory) and Intel(R) Xeon(R) Gold 6342 CPU @ 2.80GHz CPU with 18 cores (each core has 4 Gigabytes memory).\par
\begin{figure*}[h]
\centering
\includegraphics[width=0.86\linewidth]{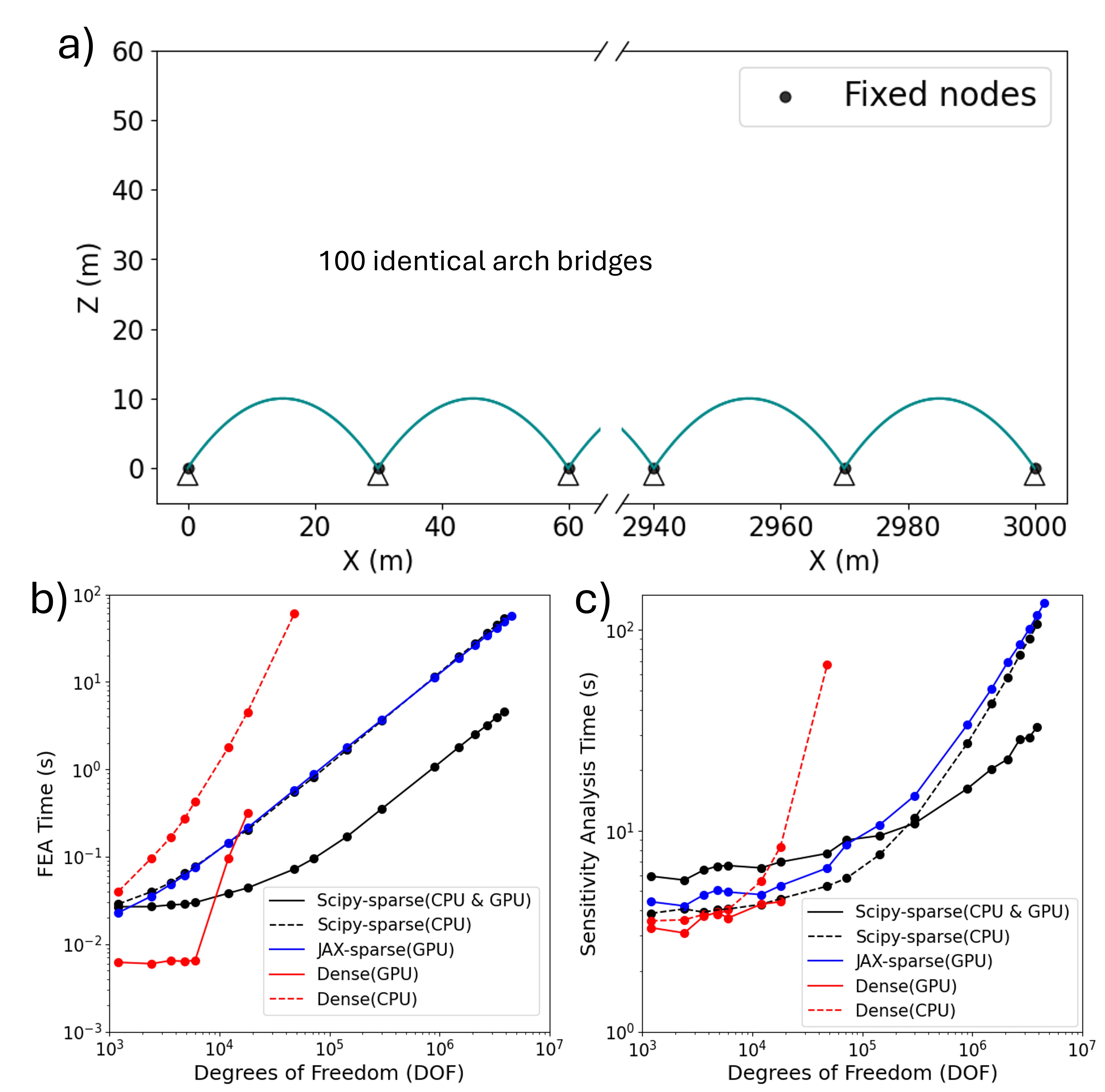}
\caption{Performance of JAX-SSO: a) Structure for performance study: 100-span arch bridge system; b) performance of JAX-SSO for FEA; c) performance of JAX-SSO for sensitivity analysis}
\label{fig:perfromance}
\end{figure*}

\begin{table}[h]
\centering 
\caption{FEA time (in seconds) of JAX-SSO for different solver options (selected)}\label{table:fea_time}

\begin{tabular}{{l}{l}{l}{l}{l}{l}}
DOF & Dense&Dense & JAX-Sparse & Scipy-Sparse  & Scipy-Sparse  \\
 & CPU & GPU & GPU & CPU & CPU\&GPU \\
\hline
1206 & 0.040 & 0.006 & 0.023 & 0.029 & 0.027\\
3606 & 0.167 & 0.006 & 0.048 & 0.051 & 0.028 \\
6006 & 0.430 & 0.007 & 0.075 & 0.077 & 0.030 \\
12006 & 1.766 & 0.095 & 0.144 & 0.143 & 0.038 \\
18006 & 4.518 & 0.314 & 0.214 & 0.203 & 0.044 \\
48006 & 60.403 & N/A & 0.580 & 0.547 & 0.072 \\
900006 &N/A & N/A & 11.154 & 11.382 & 1.061 \\
1.5 million & N/A & N/A & 18.719 & 19.512 & 1.786 \\
2.7 million & N/A & N/A& 33.702 & 36.327 & 3.184 \\
3.9 million &N/A & N/A & 48.718 & 52.845 & 4.59 \\
\hline
\end{tabular}
\end{table}
The example problem to solve is a system of 2D arch bridges consisting of beam elements (Figure \ref{fig:perfromance}.a). Consider a 100-span arch bridge system where each span is a parabolic arch spanning 30m with a rise of 10m. Distributed downward nodal load is applied to simulate the self-weight of the structure and piers are simplified as pin supports. To investigate the performance of JAX-SSO and its scalability, evenly spaced beam elements are used to discretize the structural system and the problem dimension is determined by the number of beam elements that ranges from 2 to 7500 per span. The corresponding degrees of freedom (DOF) of the whole structural system range from 1206 to about four million.\par
Figure \ref{fig:perfromance}.b and Table \ref{table:fea_time} present the FEA time of JAX-SSO and the corresponding problem dimension in terms of the number of degrees of freedom (DOF). When the problem dimension is small (DOF between 1206 and 6006), the dense solver on GPU is the fastest among all the solver options. For instance, when DOF is 6006, it takes 6.52 milliseconds for the dense solver on GPU to conduct FEA, which is about 66 times faster than the dense solver on CPU, 12 times faster than Scipy-Sparse (CPU), 5 times faster than Scipy-Sparse (CPU \& GPU) and 12 times faster than JAX-SPARSE (GPU). However, as the dimension of the problem increases, the FEA time of dense solvers on both CPU and GPU increases significantly and dense solvers become more time-consuming than all sparse solving strategies. In addition, the memory is exhausted when DOF is 48006 and 72006 for the GPU and CPU dense solvers, respectively. For sparse solvers, the performance of Scipy-Sparse (CPU) is very similar to that of JAX-SPARSE (GPU). The performance of Scipy-Sparse (CPU \& GPU) is the best where the assembly of the linear system of equations is boosted by GPU. When each arch bridge is discretized into 6500 beam elements and the DOF of the structural system is about 3.9 million, the FEA time of Scipy-Sparse (CPU \& GPU) is 4.59s, compared to 48.7s of JAX-SPARSE (GPU) and 52.8s of Scipy-Sparse (CPU). \par
The performance tests illustrate that JAX-SSO is efficient in conducting FEA. When dealing with small-scale problems, the dense solver on GPU is recommended while for large-scale problems, the Scipy-Sparse (CPU \& GPU) option is recommended. 

\subsection{Structural Optimization with JAX-SSO}
In this subsection, we introduce the formulation of structural optimization problems and how JAX-SSO efficiently addresses the challenges in structural optimization problems.
\subsubsection{Structural Optimization and Sensitivity Analysis with JAX-SSO}
Consider the following problem where we are interested in finding the optimal design parameters $\bm{p} \in \mathbb{R}^{n_P}$ of structures that minimizes the objective function $g(\bm{u},\bm{p})$ where $\bm{u}$ is solution of $\bm{Ku=f}$ that depends on the design parameters $\bm{p}$:
\begin{subequations}\label{eq:opt_formulation}
\begin{alignat}{2}
&\text{minimize} \quad \quad g(\bm{u},\bm{p}) \quad         \\
&\text{subject to: } \quad \bm{K}(\bm{p})\bm{u} =\bm{f}(\bm{p})    \quad
\end{alignat}
\end{subequations}

The gradient of the objective function $g$ with respect to the design parameters $\bm{p}$, i.e., the sensitivity, if of great interest to us because it drives gradient-based optimization. Adjoint sensitivity method is advantageous over direct sensitivity analysis when the number of design parameters is greater than the number of objective functions which is the case here and we thus implement the adjoint sensitivity \cite{Keulen2005Review}. The \textit{adjoint equation } of $\bm{Ku=f}$ reads:
\begin{equation}\label{eq:adjoint_eq}
        \bm{K}^T\bm{\lambda}=\frac{\partial g^T}{\partial \bm{u}}
\end{equation}
Where $\bm{\lambda} \in \mathbb{R}^{dof}$ is the solution to the adjoint equation, $\frac{\partial g}{\partial \bm{u}} \in \mathbb{R}^{dof}$ is the partial derivative of $g$ with respect to $\bm{u}$, which is easy to obtain since the objective function $g$ is a known function. The sensitivity, which is the gradient of the objective function with respect to the design parameters, can then be expressed in the following form:
\begin{equation}\label{eq:adjoint_sens}
        \frac{dg}{d\bm{p}} = \frac{\partial g}{\partial \bm{p}} -
        \bm{\lambda}^T(\frac{\partial \bm{K}}{\partial \bm{p}}\bm{u} - \frac{\partial \bm{f}}{\partial \bm{p}})
\end{equation}
In JAX-SSO, we define customized vector-Jacobian products (vJp) to implement the adjoint method for sensitivity analysis, similar to the approach in \cite{Pastrana2023JAX}. Because of the implementation of AD, the calculation of the sensitivity is all automatic, without the need for hand-derivation or numerical differentiation. With the sensitivity, one can leverage various gradient-based optimization algorithms to iteratively update the design parameters $\bm{p}$ to solve the structural optimization problem (Equation \ref{eq:opt_formulation}):
\begin{equation}\label{eq:update_p}
        \bm{p}^{(k+1)} = \bm{p}^{(k)} + \bm{d}^{(k)}
\end{equation}
Where the superscript $(k)$ represents the $k$-th iteration while $\bm{k}$ is the search direction based on the optimization algorithm of selection and depends on the sensitivity $\frac{dg}{d\bm{p}}$.
In JAX-SSO, the module \texttt{SSO\_model.py} is intended for sensitivity analysis and structural optimization (Figure \ref{fig:jaxsso}). Some key methods include: \texttt{add\_parameter} that adds optimization parameters, \texttt{set\_objettive} that defines the objective function to minimize, and \texttt{value\_grad\_params} that leverages AD and the adjoint method to calculate the sensitivity. Code snippet \ref{code:ssomodel} shows how to use JAX-SSO to conduct sensitivity analysis.

\begin{lstlisting}[language=Python, label={code:ssomodel}, caption={Sensitivity analysis in JAX-SSO}]
# Create the SSO model for structural optimization
from JaxSSO.SSO_model import NodeParameter,SSO_model
sso_model = SSO_model(model) # initial sso_model based on existing model
for node in design_nodes:
    nodeparameter = NodeParameter(node,2) # create a node parameter 
    sso_model.add_nodeparameter(nodeparameter)  # add node parameter to sso_model

#Initial the parameters and set objective function
sso_model.initialize_parameters_values()
sso_model.set_objective(objective='strain energy',func=None,func_args=None) 

#Conduct sensitivity analysis
sso_model.value_grad_params(which_solver='sparse',enforce_scipy_sparse = True)
\end{lstlisting}

\subsubsection{Validation of JAX-SSO for sensitivity analysis}
\begin{figure*}[h]
\centering
\includegraphics[width=1\linewidth]{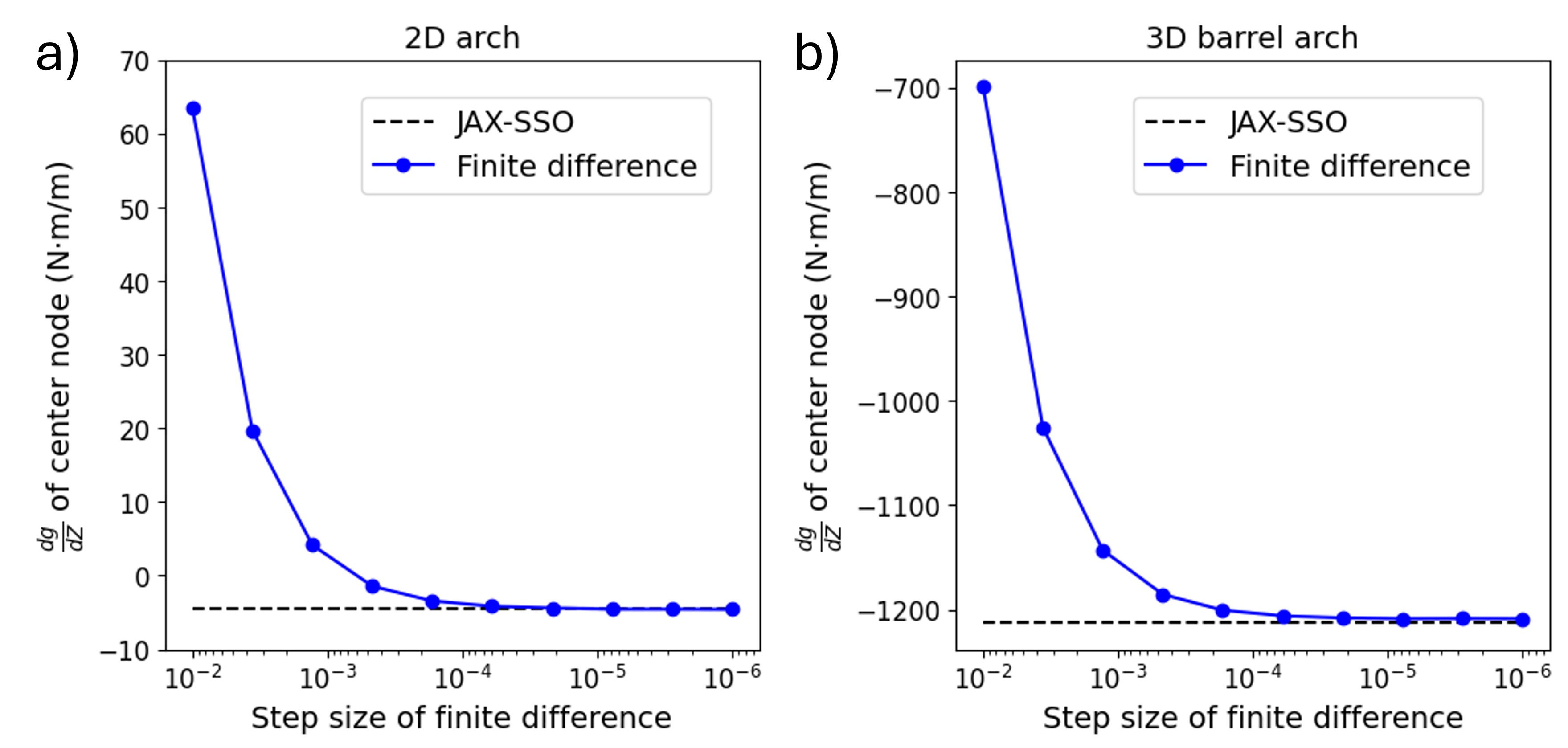}
\caption{Validation of JAX-SSO for sensitivity analysis: a) center node of 2D arch; b) center node of 3D barrel arch}
\label{fig:ad_valid}
\end{figure*}
We compare the sensitivity evaluation from JAX-SSO against finite difference to validate the accuracy of JAX-SSO in sensitivity calculations. Same structures as the ones presented in the FEA-validation section (Figure \ref{fig:fea_valid}) are used.\par
For both structures, the design parameter $\bm{p}$ is set as the Z-coordinates of the nodes and the objective function $g$ is set as total strain energy of the system, $g=0.5\bm{f}^T\bm{u}$. Figure \ref{fig:ad_valid} shows the values of $\frac{dg}{d\bm{p}}$ from JAX-SSO and finite difference with different step sizes: here we present the derivative of $g$ with respect to the Z-coordinate of the center node from each structural system for illustration purposes. For the 2D arch with beam elements, JAX-SSO outputs -4.54617 N$\cdot$m/m while the finite difference method outputs -4.56982 N$\cdot$m/m when the step size is $10^{-6}$, showing a difference of 0.5$\%$. Similarly, for the 3D barrel arch with shell elements, -1212.6236 N$\cdot$m/m is obtained by JAX-SSO while finite difference gives -1208.8802 N$\cdot$m/m when the step size is $10^{-6}$, which is a difference of 0.3$\%$. The results validate the accuracy of JAX-SSO for sensitivity analysis.

\subsubsection{Performance of JAX-SSO for sensitivity analysis}
The performance of JAX-SSO is evaluated in terms of its speed in conducting sensitivity analysis. The tests are conducted using the same hardware as in Section \ref{sec:fea_per}. The structure to analyze (Figure \ref{fig:perfromance}.a)  and the FEA options (Table \ref{table:fea_options}) are the same as the ones used in Section \ref{sec:fea_per} where we present the performance of JAX-SSO for FEA. The objective function $g$ is the total strain energy of the system $g=0.5\bm{f}^T\bm{u}$ and the design parameters $\bm{p}$ are the nodal Z-coordinates of every node in the system.\par
Figure \ref{fig:perfromance}.c presents the time needed to conduct sensitivity analysis using JAX-SSO with different solving options. The trend is very similar to Figure \ref{fig:perfromance}.b where we present the FEA time of JAX-SSO. When the problem dimension is small (less than 104 DOF), dense solvers (on both CPU and GPU) work better than sparse solvers in conducting sensitivity analysis. However, when the problem dimension increases, the performance of dense solvers either is worse than sparse solvers or runs out of memory. For instance, when DOF is 48006, dense solver with GPU fails to output the sensitivity $\frac{dg}{d\bm{p}}\in \mathbb{R}^{8001}$ due to insufficient memory; dense solver with CPU is able to output the sensitivity but it takes around 67 seconds, which is much slower than all the sparse solvers. When the DOF of the problem increases to more than 106, the Scipy-Sparse (CPU \& GPU) option outperforms all the other sparse solvers. For instance, when DOF is around 3.9 million, it takes Scipy-Sparse (CPU \& GPU) 32.9 seconds to conduct the sensitivity analysis, compared to 118.3 seconds for JAX-SPARSE (GPU) and 106.6 seconds for Scipy-Sparse (CPU \& GPU). When dealing with small-scale problems, the dense solver on GPU is recommended while for large-scale problems, the Scipy-Sparse (CPU \& GPU) option should be used. 

\subsection{Integration with Neural Networks using JAX-SSO for Structural Optimization}\label{sec:NN_FLOW_1}
\begin{figure*}[h]
\centering
\includegraphics[width=1\linewidth]{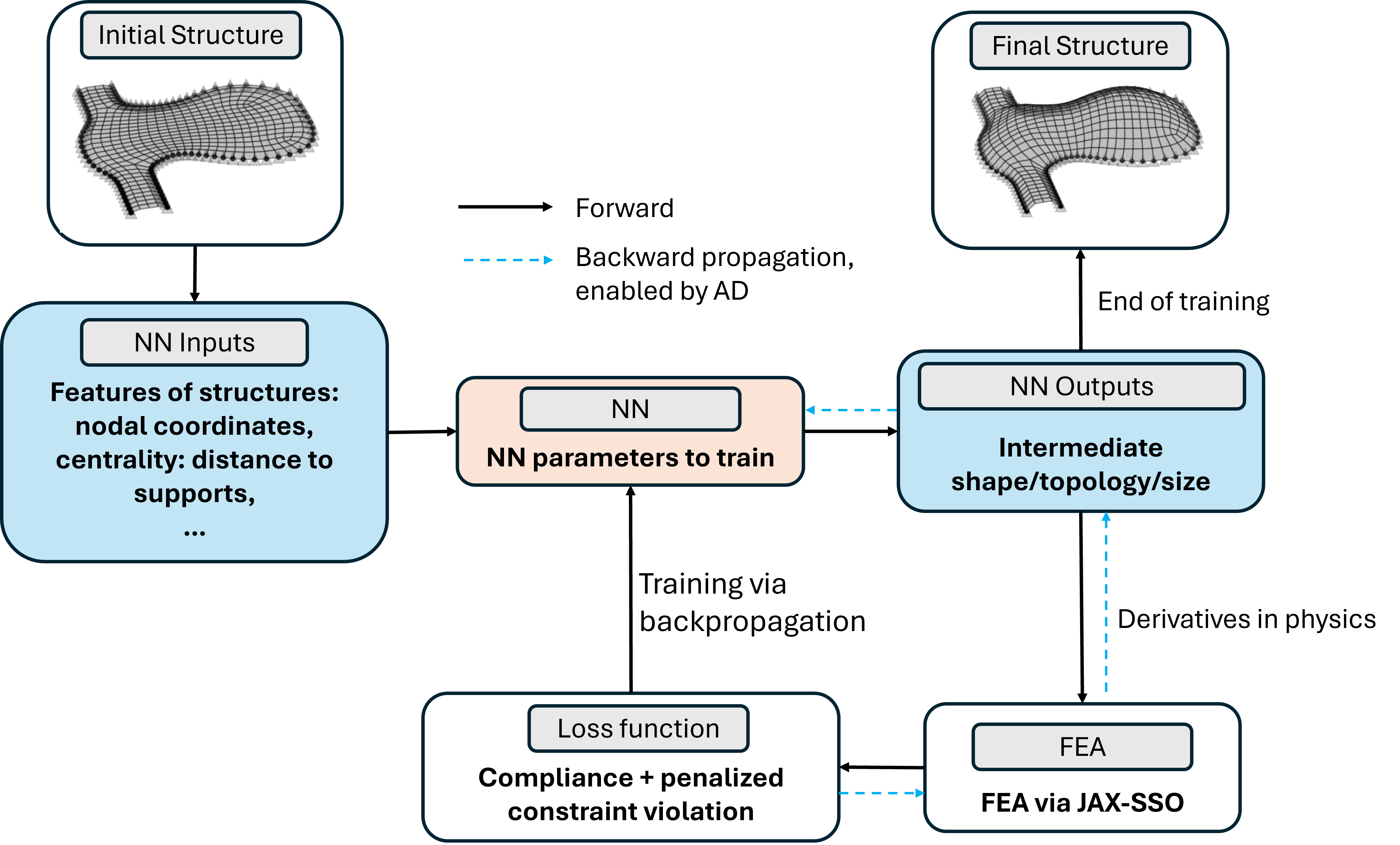}
\caption{Structural optimization via NN reparameterization}
\label{fig:nn_flow_1}
\end{figure*}
Written in Python, JAX-SSO naturally supports seamless integration with various machine learning (ML) Python libraries such as JAX, TensorFlow and PyTorch. As both JAX-SSO and ML libraries operate on the same platform, the efforts spent on transporting data from traditional FEM solver to Python can be saved, which facilitates more efficient data preparation and training. Another advantage brought by the AD feature in JAX-SSO is that derivatives from physics can be easily chained together with the derivatives in Neural Networks (NN) to enable the training of physics-informed NN. Here we present how JAX-SSO can be integrated with NN to assist structural optimization problems.\par
The application of physics-informed NN has emerged in the context of structural optimization in recent years and one approach is to use NN to parametrize optimization problems and to solve the structural optimization problems through the training of NN. Hoyer et al. \cite{Hoyer2019Neural} shows that by parametrizing the topology optimization problems using convolutional neural networks (CNN) and training a physics-informed CNN, the optimization result outperforms optimization methods with only physics. Chandrasekhar and Suresh \cite{Chandrasekhar2021TOuNN:} implemented NN to conduct topology optimization. Favilli et al. \cite{Favilli2024Geometric} proposed a method that implements geometric deep learning to reparametrize shape optimization problems to embed design intent implicitly in the training of NN so that the optimization result preserves specific geometric features.\par
We use the framework illustrated in Figure \ref{fig:nn_flow_1} to incorporate JAX-SSO with NN for structural optimization problems. Taking shape optimization as an example, for an initial structure with series of nodes with pairs of (X,Y) coordinates, we want to find a series of Z coordinates that minimize the objective function. Instead of directly using gradient-based optimization algorithm to iteratively update Z, we train a NN that outputs Z given a pair of input (X,Y) for each node. One first initializes NN and inputs the NN with per-node features, which are the (X,Y) coordinates in this case. The NN then outputs Z and one can conduct FEA using JAX-SSO. The loss function to train the NN consists of the objective function one expects to minimize (such as the compliance of the structure) and some penalty terms for constraints violation, if any. The next step is to update NN parameters using backpropagation (dashed blue line in Figure \ref{fig:nn_flow_1}) and optimization method of selection (for instance, Adam algorithm \cite{Kingma2017Adam:}): taking the derivatives of the loss function with respect to parameters in NN, such as the weights and bias. Due to the AD feature in JAX-SSO, the derivatives within the FEA domain can be seamlessly linked to the derivatives in the NN domain through the chain rule. After backpropagation, the NN parameters are updated, and the next iteration starts. This training process of NN continues iteratively until meeting the stopping criterion one specifies and the final structure is found. \par 
The employment of NN in this approach differs from other NN applications in structural optimization where NN is implemented as a surrogate or acceleration to gradient-based optimization methods. Here NN is implemented for the reparameterization of structural optimization problems: through adding another NN domain to the optimization problem, the design space is reparametrized and spanned by NN parameters.

\section{Examples}\label{sec:ex}

\subsection{Shape optimization of grid shell: Station}
\begin{figure*}[h]
\centering
\includegraphics[width=1\linewidth]{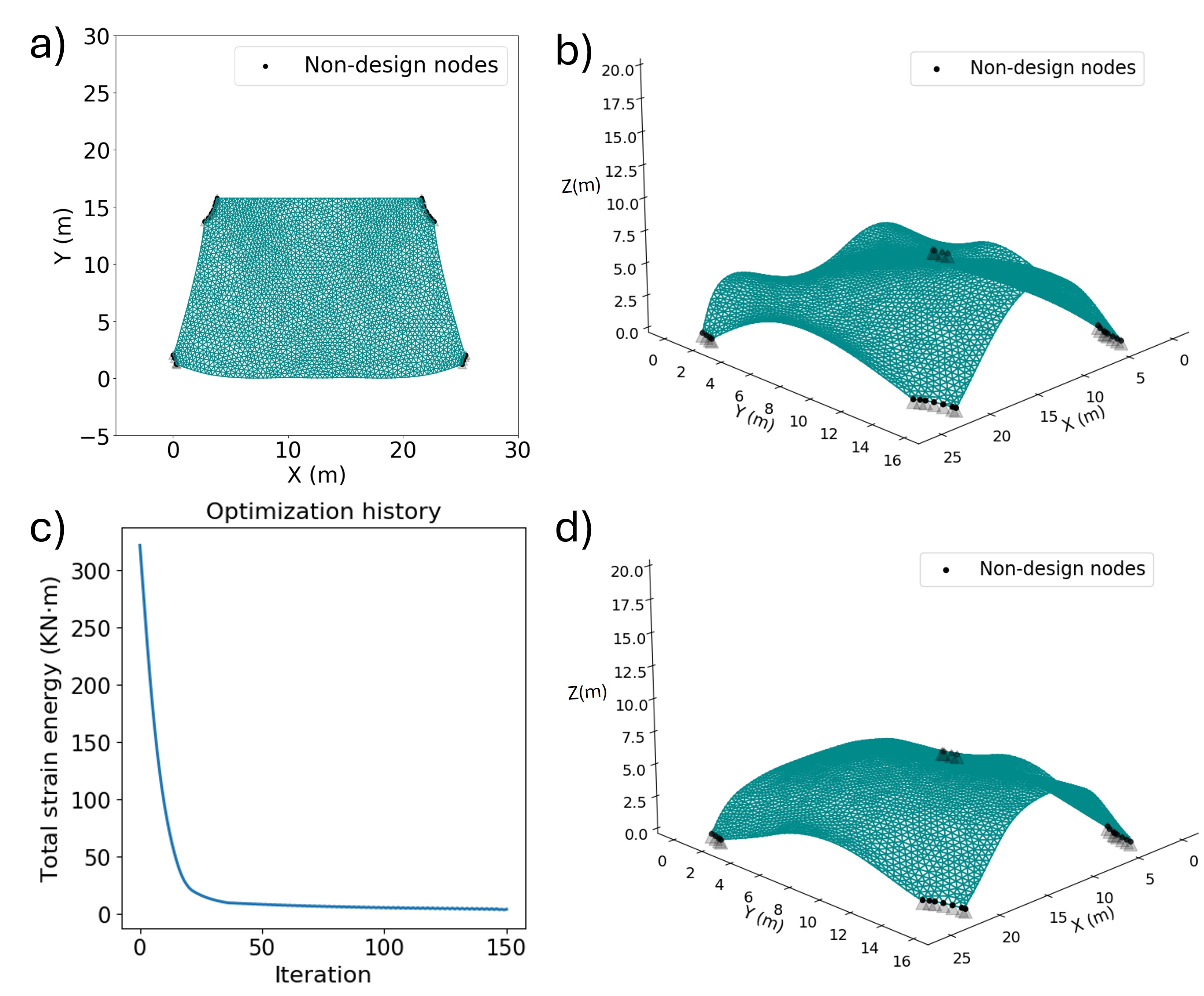}
\caption{Shape optimization of Station: a) Plan view; b) Initial structure; c) Optimization history; d) Optimized structure}
\label{fig:station}
\end{figure*}

The first example is a grid shell structure consisting of beam elements. The initial structure is shown in Figure \ref{fig:station}.a\&b, and the geometry is adapted from an example in the work by Favilli et al. \cite{Favilli2024Geometric}. The grid shell spans approximately 25 m along the X-axis and 15 m along the Y-axis. It consists of 2829 nodes and 8266 beam elements. All the elements have the same properties: Young’s modulus $E$=3790 Gpa, Poisson’s ratio $\nu$=0.3, and a rectangular cross-section with 0.2 m depth and 0.1 m width. The corner nodes are pin-supported and downward (-Z) nodal load with a magnitude of 10 KN is applied to every non-supported node.\par
The objective function g is set as the total strain energy of the system $g=0.5\bm{f}^T\bm{u}$, and the design parameters $\bm{p}$ are the nodal Z-coordinates of all the non-supported nodes. To avoid non-smooth shapes or jagged surfaces encountered by directly using the sensitivity in the optimization procedure, we apply a standard hat filter with a radius of 4m to the sensitivity $\frac{dg}{d\bm{p}}\in \mathbb{R}^{2806}$. For a complete introduction of sensitivity filtering, please refer to the work by Bletzinger \cite{Bletzinger2014consistent}. Gradient descent with a step size of 0.1 is used and maximum iteration is set as 150.\par
Figure \ref{fig:station}.b presents the optimization history in terms of the objective function value. The strain energy of the initial structure is 322.03 KN$\cdot$m and it is reduced to 4.08 KN$\cdot$m after 150 iterations, showing a 98.7$\%$ decrease in the objective function value. The optimized structure is plotted in Figure \ref{fig:station}.d. Compared to the initial structure where it presents both positive and negative Gaussian curvature, regions with positive Gaussian curvature dominate the optimized structure, which implies the structural elements mainly resist external loads by axial behavior.
\subsection{Shape optimization of continuous shell: Exhibition Hall}\label{sec:shape_ex}
\begin{figure*}[h]
\centering
\includegraphics[width=1\linewidth]{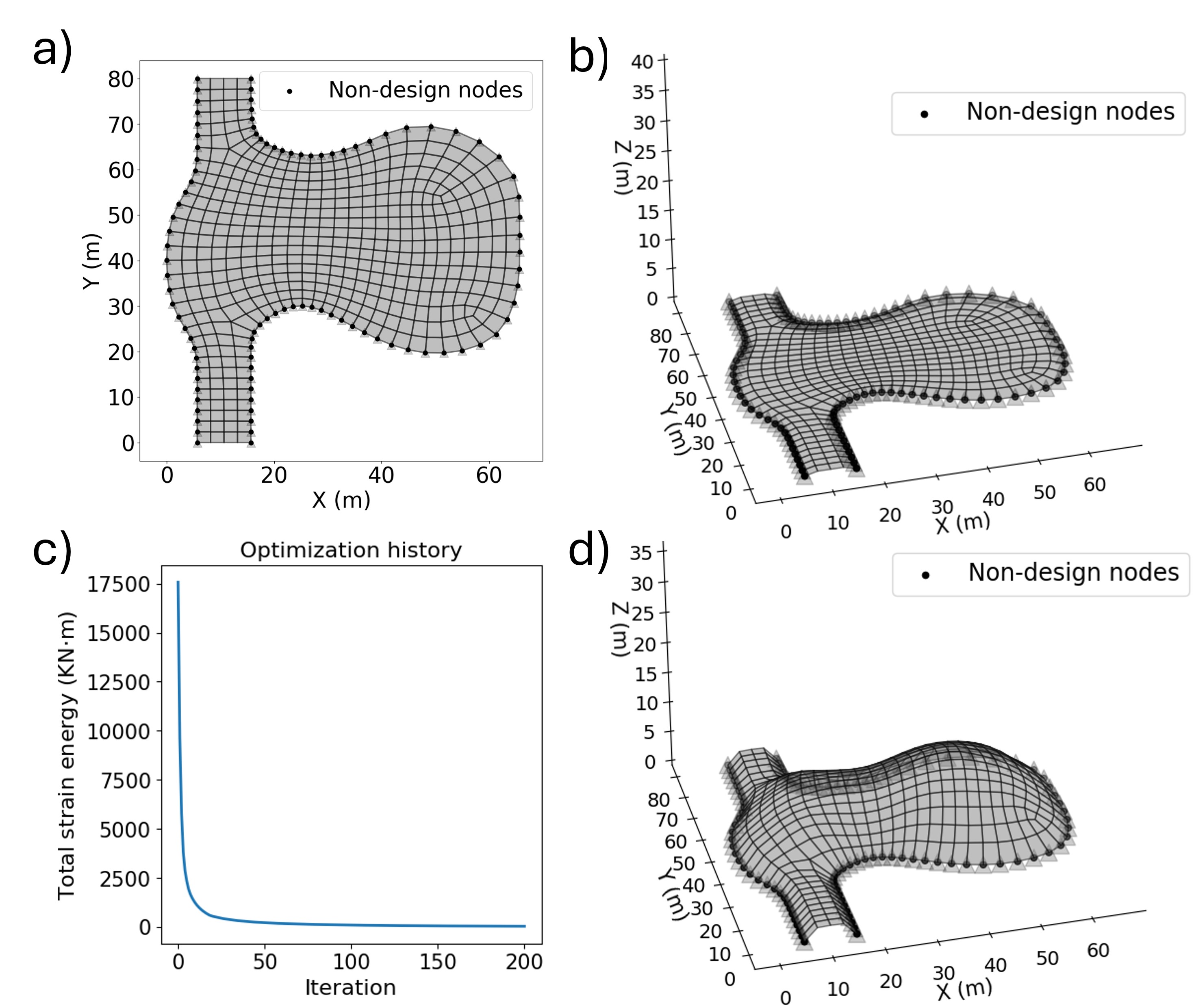}
\caption{Shape optimization of Exhibition Hall (continuous shell): a) Plan view; b) Initial structure; c) Optimization history; d) Optimized structure}
\label{fig:shape_exhi}
\end{figure*}
The second example is shape optimization of Exhibition Hall (Figure \ref{fig:shape_exhi}) that consists of quadrilateral shell elements. The plan view (Figure \ref{fig:shape_exhi}.a) is adapted from Frei Otto’s Mannheim Multihalle. The Exhibition Hall spans around 60 m and 80 m along the X-axis and the Y-axis, respectively. 457 quadrilateral shell elements are used to discretize the continuous structure. Same properties are assigned to all the shell elements: thickness is set as 150 millimeters, Young’s modulus $E$=10000 Gpa, and Poisson’s ratio $\nu$=0.3. There are 510 nodes in the system where 98 nodes are non-design nodes and are fixed. Nodal load with a magnitude of 50KN is applied to every non-supported node.\par
The shape optimization of the Exhibition Hall is based on strain energy minimization and the design variables $\bm{p}\in \mathbb{R}^{412}$ are set as the Z-coordinates of all the non-supported nodes. To avoid significant distortion of shell elements and non-smooth shape, we apply a hat filter with a radius of 10 m to the sensitivity, just like the first example. The optimizer is gradient descent with a step size of 0.1 and the number of iterations is set as 200. \par
The design nodes in the initial structure (Figure \ref{fig:shape_exhi}.a) have randomly generated Z-coordinates between 0.5 m and 0.51m. The initial structure acts like a giant flat slab that resists load mainly by bending behavior, which is not the most efficient load-bearing mechanism for continuous shells. The initial structure thus has a large strain energy value: 17573.4 KN$\cdot$m. After 200 iterations, the total strain energy is reduced to 28.9 KN$\cdot$m, showing a decrease of 99.8$\%$ (Figure \ref{fig:shape_exhi}.c). The optimized structure is plotted in Figure \ref{fig:shape_exhi}.d: at Y=0 m and Y=80 m, there are two openings to the exhibition area through two-barrel arches; between Y=20m and Y=60m is the main exhibition area where a doubly curved shell can be observed; around X=25m, negative Gaussian curvature can be seen whereas when X$<$20m or when X$>$30m, a dome-like shape is found.
\subsection{Size optimization of continuous shell: Exhibition Hall}
In this example, we leverage JAX-SSO to conduct size optimization for a continuous shell, Exhibition Hall. The initial structure to optimize is the shape-optimized Exhibition Hall from Section \ref{sec:shape_ex}, which is illustrated in Figure \ref{fig:shape_exhi}.d. The cross-sectional properties of the shell elements are the same as Section \ref{sec:shape_ex} where all the shell elements have the same thickness of 150mm. 
We follow Hasançebi’s work \cite{Hasançebi2008Adaptive} to formulate the size optimization problem to minimize the total material consumption while preserving expected properties of the structure. Here we limit the maximum nodal displacement and constrain the minimum thickness of the shell structure:
\begin{subequations}\label{eq:size_formulation}
\begin{alignat}{2}
&\text{minimize} \quad \quad g(\bm{p}) = (1+\epsilon_1c(\bm{p}))W(\bm{p}) \quad         \\
&\text{subject to: } \quad \bm{K}(\bm{p})\bm{u} =\bm{f}    \quad
\end{alignat}
\end{subequations}
Where the design parameters $\bm{p}\in \mathbb{R}^{417}$ are the shell elements’ thickness values for all the 417 quadrilateral shell elements and the objective function $g(\bm{p})$ is the penalized total material volume of the structure. The first term of $g(\bm{p})$, $(1+\epsilon_1c(\bm{p}))$, is the penalty term, consisting of a self-adaptive coefficient $\epsilon_1$, and a term to penalize the violation of expected properties $c(\bm{p})$. Herein, we define $c(\bm{p})$ as the violation of nodal displacement:
\begin{equation}
    c(\bm{p})=\Sigma_{j=1}^{n_{node}}\max(0,\frac{|u_j|}{u_{max}}-1)
\end{equation}
Where $n_{node}$ is the total number of nodes in the system, $|u_j|$ is the absolute value of the nodal displacement of node $j$, and $u_{max}$ is the maximum allowed nodal displacement. If one node violates the displacement requirement, it adds up to the penalty $c(\bm{p})$. Coefficient $\epsilon_1$ is self-adaptive depending on if there is violation of the expected properties:
\begin{equation}
    \epsilon_1^{(k+1)} = 
    \begin{cases}
      (\frac{1}{\kappa})\epsilon_1^{(k)}, & \text{if no violation}\ a=1 \\
      \kappa\epsilon_1^{(k)}, & \text{if violation}
    \end{cases}
\end{equation}
Where $\kappa$ is the learning parameter of $\epsilon_1$, and it is set as 1.015 for this example; the superscript $(k)$ denotes the $k$-th iteration in the optimization. The second term of $g(\bm{p})$, $W(\bm{p})$, is the material volume with penalty if there is minimum thickness violation:
\begin{equation}
    W(\bm{p})=\Sigma_{i=1}^{n_{quad}}A_ip_i+\Sigma_{i=1}^{n_{quad}}\max(0,\frac{t_{min}-p_i}{t_{min}})
\end{equation}
Where $n_{quad}$ is the number of quadrilateral shell elements, $A_i$ is the surface area of the $i$-th element, $p_i$ is the $i$-th design parameter (the thickness), and $t_{min}$ is the minimum allowable thickness. When the design thickness is less than the requirement $t_{min}$, it penalizes the total volume. In this example, $t_{min}$ is set as 50mm and the maximum allowable nodal displacement $u_{max}$ is set to be 1.5 cm which is applied to all the nodal displacement along the Z-direction.\par
\begin{figure*}[h]
\centering
\includegraphics[width=1\linewidth]{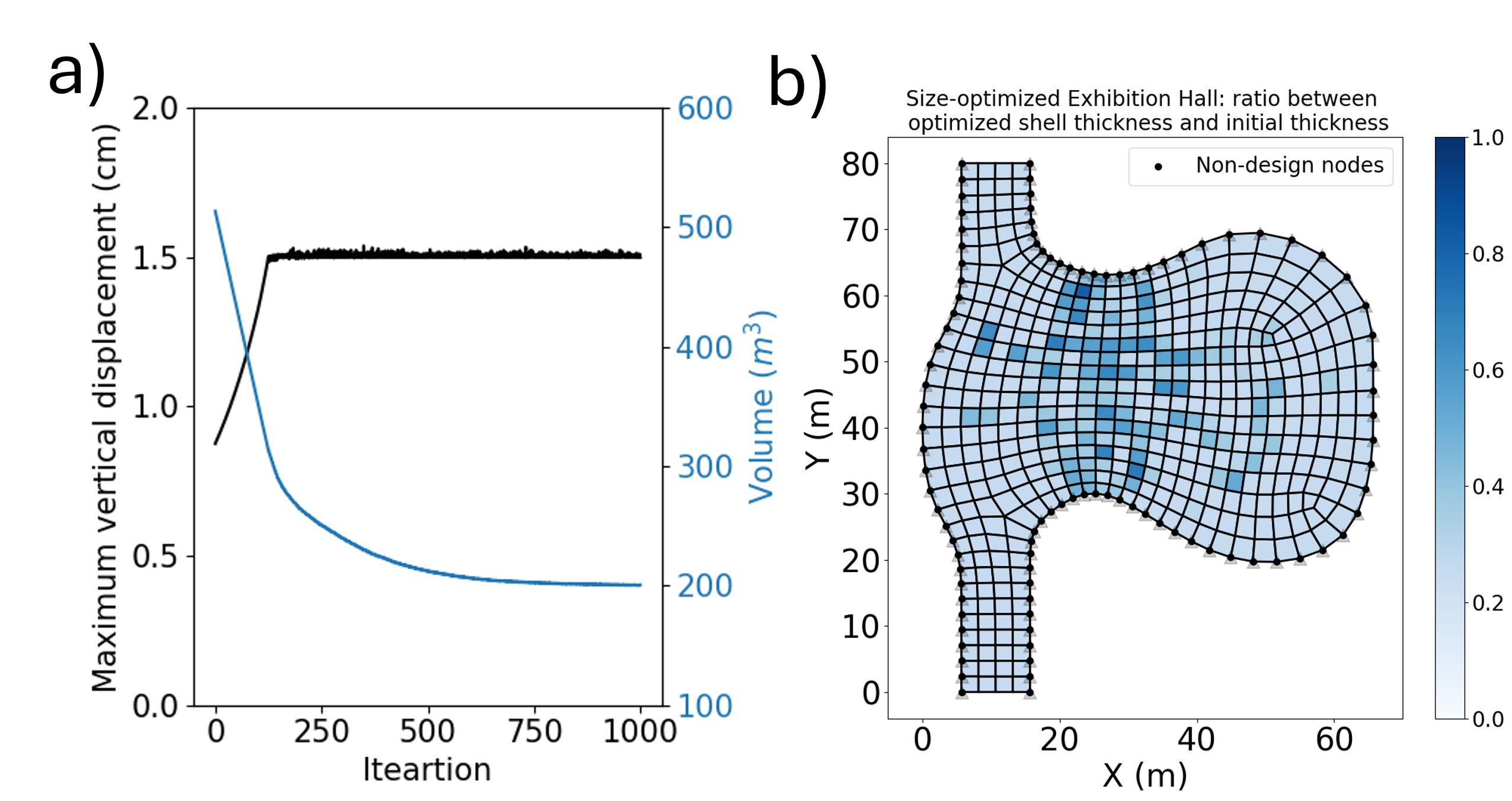}
\caption{Size optimization of Exhibition Hall (continuous shell): a) Optimization history; b) Plan view of the optimized structure where the color represents the ratio between the optimized shell-element thickness and initial shell-element thickness (150mm) }
\label{fig:size_exhi}
\end{figure*}
The sensitivity of the size optimization problem,  $\frac{dg}{d\bm{p}}$, is obtained easily via JAX-SSO thanks to its AD feature. We directly apply Gradient Descent with a step size of 0.001 and the number of iterations is set as 1000. Figure \ref{fig:size_exhi}.a illustrates the volume change and change of the maximum nodal Z-displacement throughout the optimization history. The total volume of the initial unoptimized shell structure is 513.4 m$^3$ and it is reduced to 200.6 m$^3$ after 1000 iterations, showing a decrease of 60.9$\%$. Meanwhile, the maximum nodal Z-displacement increases from 0.88cm to 1.50cm after 1000 iterations due to reduced stiffness led by thinner shell elements. Due to the penalty term $c(\bm{p})$ added to the objective function $g$, the maximum nodal Z-displacement does not increase continuously after it first reaches the prescribed limit 1.5cm at the 125th iteration. Figure \ref{fig:size_exhi}.b illustrates the result of the size optimization for the Exhibition Hall in terms of the ratio between the optimized thickness of shell elements and the initial thickness of 150mm. The shell elements around X=25m tend to have thicker cross-sections where the structure presents negative Gaussian curvature as shown in Figure \ref{fig:shape_exhi}.d. A possible explanation is that, due to the negative Gaussian curvature, the shell elements around X=25m do not carry external loads using the material efficient membrane behavior, but rather by bending behavior, which is material inefficient for shells. On the contrary, most quadrilateral shell elements have around the minimum allowable thickness $t_{min}$ of 50mm when they are away from X=25m, where the shell structure either presents dome-like shape or barrel arch-like shape, which are both membrane-force dominant shapes, explaining why $t_{min}$ is enough to satisfy the maximum nodal displacement requirement.

\subsection{Simultaneous shape and topology optimization of continuous shell: case 1} \label{sec:topo_shape_1}
JAX-SSO can be leveraged to solve topology optimization problems of shells. We consider a more generalized example herein where we conduct shape and topology optimization simultaneously. The design variables $\bm{p}$ has two components: $\bm{p}_S$ as shape design variables and $\bm{p}_T$ for topology. Here the shape variables are normalized Z-coordinates of nodes where the $i$-th shape variable reads:
\begin{equation}\label{eq:nomarl_z}
    p_{S,i}=(Z_i-Z_{min})/(Z_{max}-Z_{min})
\end{equation}
\begin{figure*}[h]
\centering
\includegraphics[width=1\linewidth]{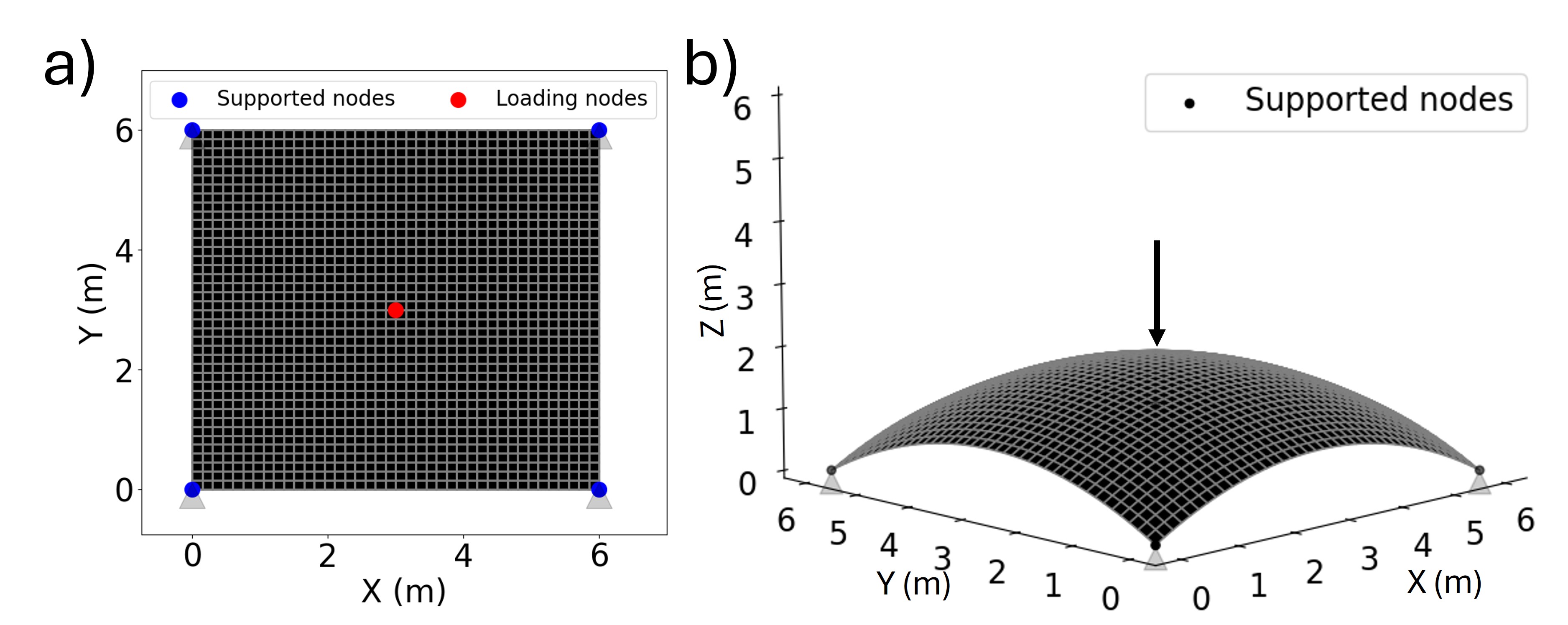}
\caption{Initial structure for simultaneous shape and topology optimization, case 1: a) plan view; b) perspective view}
\label{fig:topo_ini}
\end{figure*}
\begin{figure*}[h]
\centering
\includegraphics[width=0.95\linewidth]{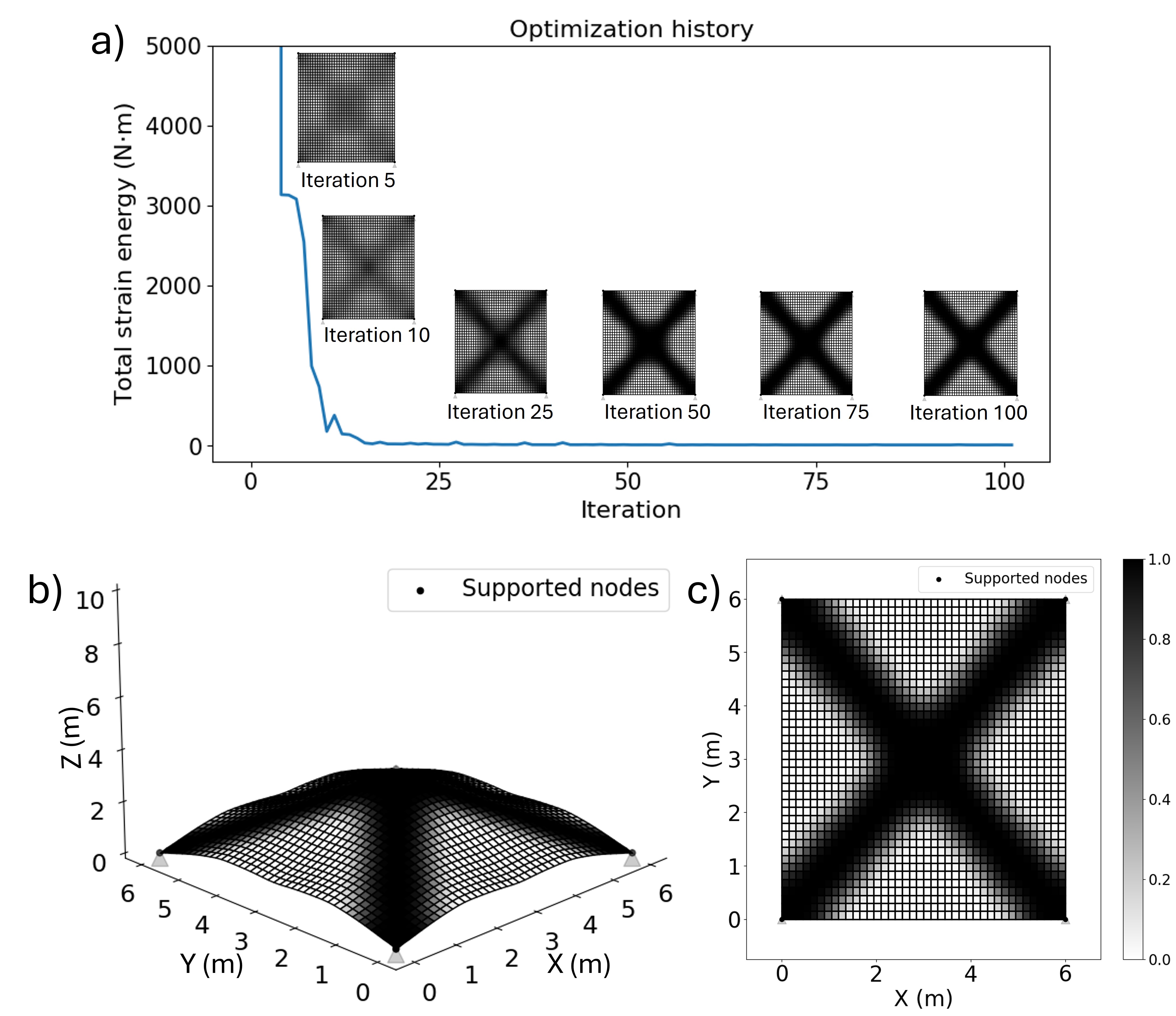}
\caption{Case 1 of simultaneous shape and topology optimization: a) Optimization history; b) Final structure: perspective view; c) Final structure: plan view with effective density distribution graph where white regions represent }
\label{fig:topo_final}
\end{figure*}
where $Z_i$ is the nodal Z-coordinate of the $i$-th design node, $Z_{max}$ and $Z_{min}$ are the upper and lower bounds for the Z-coordinate, respectively. In terms of topology optimization, we adopt the Solid Isotropic Material with Penalization (SIMP) method \cite{Bendsoe2013Topology} and thus, the topology variable $\bm{p}_T$ is the effective density ratio for each quadrilateral shell element. For the $j$-th shell element, the Young’s modulus $E_j$ is adjusted by the following according to SIMP method:
\begin{equation} \label{eq:simp_eq}
    E_j=(p_{T,j})^PE_{*.j}
\end{equation}
where $P$ is the penalty factor used to suppress intermediate values of the topology variables, $E_{*,j}$ is the referenced original Young’s modulus for the $j$-th element, and the $j$-th topology variable $p_{T,j}$ is limited between a lower bound $p_{T,min}$ to avoid singularity in the global stiffness matrix and 1: $p_{T,min}\leq p_{T,j}\leq 1$.\par
We consider a classic problem: strain energy minimization of simply supported shells. The initial structure is illustrated in Figure \ref{fig:topo_ini}, where it spans 6 m along both X and Y axes with a rise of 1.8 m. It is discretized into 40 by 40 quadrilateral shell elements and there are 1681 nodes in the structural system. The four corner nodes are simply supported, and a downward point load of 500KN is applied to the center of the shell structure. All the shell elements have the same sectional properties: thickness is set as 150 millimeters, reference Young’s modulus $E_*$=20000 Gpa, and Poisson’s ratio $\nu$=0.3.\par
The objective function $g$ is set as the total strain energy of the system, $g=0.5\bm{f}^T \bm{u}$. The design variables $\bm{p}$ are made up of normalized nodal Z-coordinates of all the 1677 non-supported nodes as shape variables $\bm{p}_S \in \mathbb{R}^{1677}$ and the effective density ratio values of all the 1600 shell elements $\bm{p}_T \in \mathbb{R}^{1600}$ as topology variables. We apply a standard hat filter with radius of 1.5m to the shape variables $\bm{p}_S$  and a hat filter with radius 0.5m to the topology variables $\bm{p}_T$. The sensitivity $\frac{dg}{d\bm{p}} \in \mathbb{R}^{3277}$ is automatically calculated by JAX-SSO. We impose constraints on all the design variables: $0.01<\bm{p}<1$ and set the upper and lower bounds for Z-coordinates to $Z_{max}=3$m and $Z_{min}=0$m. In addition, we limit the total material volume usage to around 50$\%$ of the initial continuous shell structure via a material volume constraint: $\Sigma_{j=1}^{j=1600} p_{T,j}<800$. The penalty factor $P$ for the Young’s modulus modification is set as 7 (Equation \ref{eq:simp_eq}). The optimization algorithm used is the Method of Moving Asymptotes (MMA) \cite{Svanberg1987method} and the maximum iteration number is set as 100. The initial topology variable is set as 0.1 for all $\bm{p}_T$.\par
Figure \ref{fig:topo_final}.a presents the optimization history of the total strain energy. After 100 iterations, the strain energy is reduced from to 2.8$\times10^{10}$ N$\cdot$m to 4.72 N$\cdot$m by conducting simultaneous shape and topology optimization. During the optimization history (Figure \ref{fig:topo_final}.a), the effective density is gradually increased in certain areas: areas around the center of the structure where the load is applied and areas connecting the center to the supports. The final optimized structure is shown in Figure \ref{fig:topo_final}.b \& Figure \ref{fig:topo_final}.c. The final structure is a truss-like shape where the elements between adjacent supports become void as the load is now efficiently carried by elements connecting the center to the supports through mainly axial behavior. The total material volume constraint is also successfully satisfied with $\Sigma_{j=1}^{j=1600}p_{T,j}=799.9<800$.
\subsection{Simultaneous shape and topology optimization of continuous shell: case 2}
\begin{figure*}[h]
\centering
\includegraphics[width=1\linewidth]{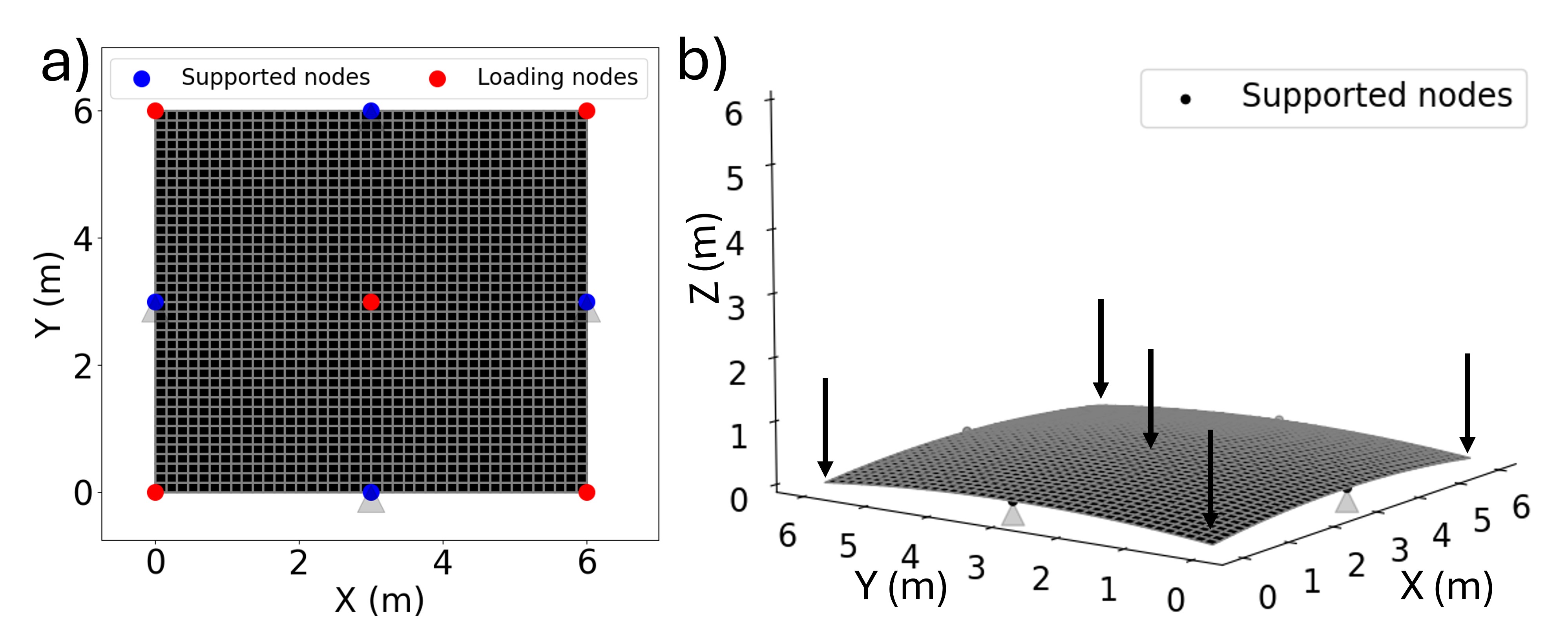}
\caption{Initial structure for simultaneous shape and topology optimization, case 2: a) plan view; b) perspective view}
\label{fig:topo_ini_2}
\end{figure*}
The same optimization scheme as Section \ref{sec:topo_shape_1} is implemented for another example on simultaneous shape and topology optimization using JAX-SSO. The initial structure is illustrated in Figure \ref{fig:topo_ini_2}. The plan view of the initial structure and the properties of shell elements are the same as the example in Section \ref{sec:topo_shape_1} whereas the loads and boundary conditions are different. 500KN downward point loads are applied to the four corner nodes and the center node, and the four nodes at the midpoint of each edge are simply supported.\par
We set the objective function as the total strain energy of the system $g=0.5\bm{f}^T \bm{u}$, and the design variables $\bm{p} \in \mathbb{R}^{3281}$ consist of normalized nodal Z-coordinates of all the 1681 nodes (including the supported nodes) for shape optimization as well as the effective density ratio values of all the 1600 shell elements for topology optimization. A standard hat filter with radius 1.5m to the shape variables $\bm{p}_S$ and a hat filter with radius 0.25m to the topology variables $\bm{p}_T$. The sensitivity $\frac{dg}{d\bm{p}}\in \mathbb{R}^{3281}$ is automatically calculated by JAX-SSO. The constraints applied to the design variables and the penalty factor $P$ for the Young’s modulus modification are the same as Section \ref{sec:topo_shape_1}. MMA is set as the optimizer and the maximum iteration number is set to be 300. The initial topology variable value is set as 0.1 for all $\bm{p}_T$.\par
\begin{figure*}[h]
\centering
\includegraphics[width=1\linewidth]{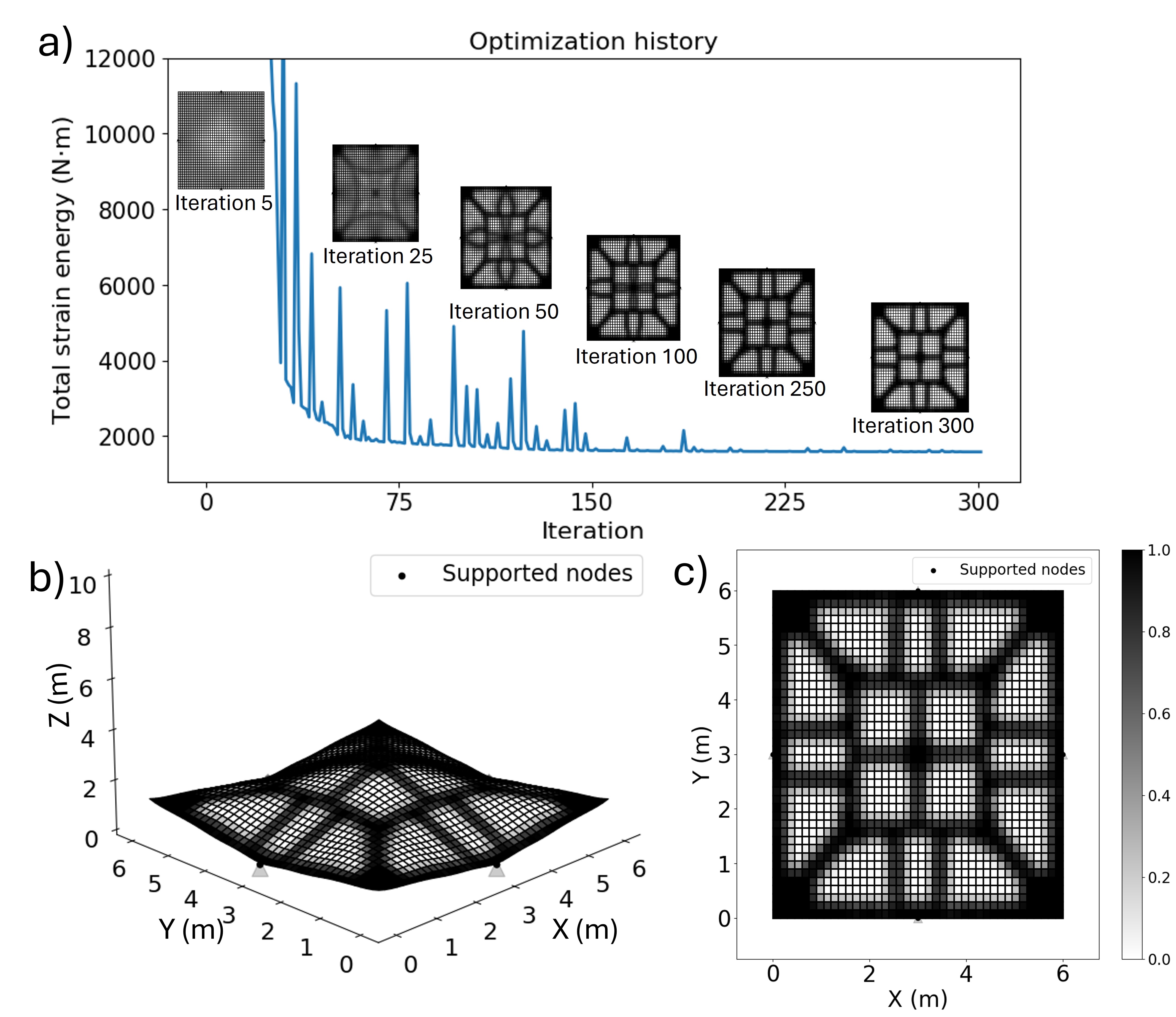}
\caption{Initial structure for simultaneous shape and topology optimization, case 2: a) plan view; b) perspective view}
\label{fig:topo_fin_2}
\end{figure*}
Figure \ref{fig:topo_fin_2}.a presents the time history of the total strain energy: within the first 200 iterations, the total strain energy has a decreasing trend while there are spikes, indicating the MMA algorithm is adjusting temporary design parameters to satisfy the constraints we impose so that the objective function value is temporarily sacrificed. After around 200 iterations, the total strain energy gradually converges to 1585.6 N$\cdot$m at 300-th iteration while no significant spike is observed.\par
The final structure is plotted in Figure \ref{fig:topo_fin_2}.b \& Figure \ref{fig:topo_fin_2}.c. From the support to the adjacent corner, the structure gradually rises upwards, and the structure presents a cantilever shape to resist the corner point load. In terms of the topology of the structure, the four corners, as well as the center are filled with shell elements to support the point loads. A two-ring pattern is observed where the inner ring transfers the center point load to a squared-shaped inner ring. Each edge of the inner square ring is connected to the outer ring through four spines: two spines leading to the support and the other two lead to the corners. The total material volume constraint is successfully satisfied with $\Sigma_{j=1}^{j=1600}p_{T,j}=799.9<800$.
\subsection{Integration with Neural Networks for Structural Optimization: simultaneous shape and topology optimization}
\begin{figure*}[h]
\centering
\includegraphics[width=1\linewidth]{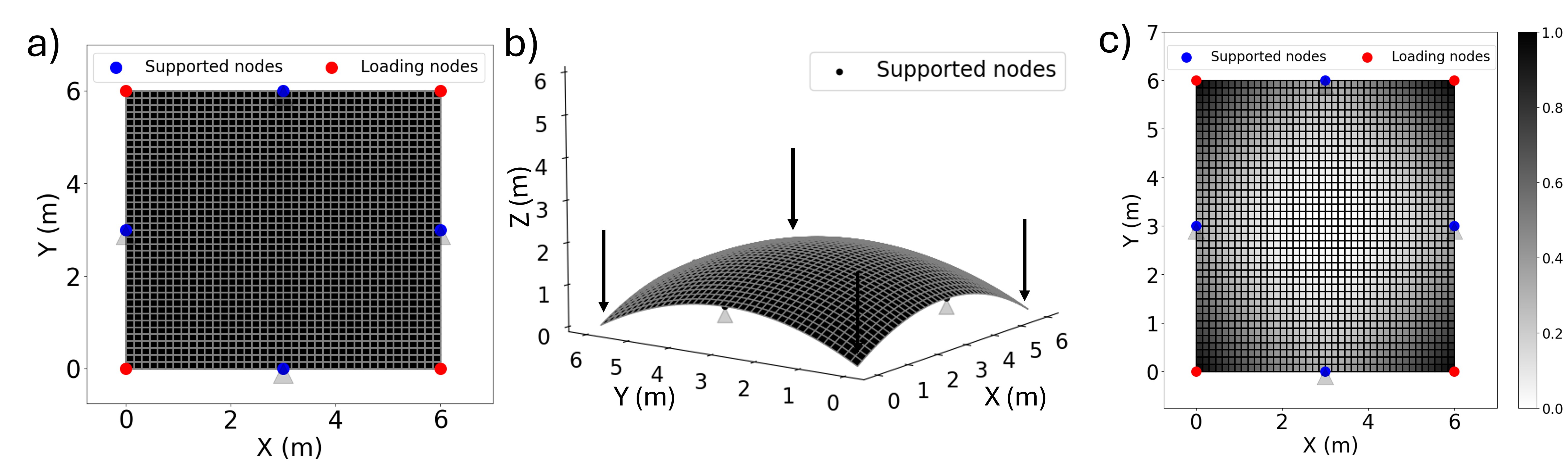}
\caption{NN for structural optimization, initial structure: a) plan view; b) perspective view; c) normalized centrality }
\label{fig:nn_ini}
\end{figure*}
\begin{figure*}[h]
\centering
\includegraphics[width=1\linewidth]{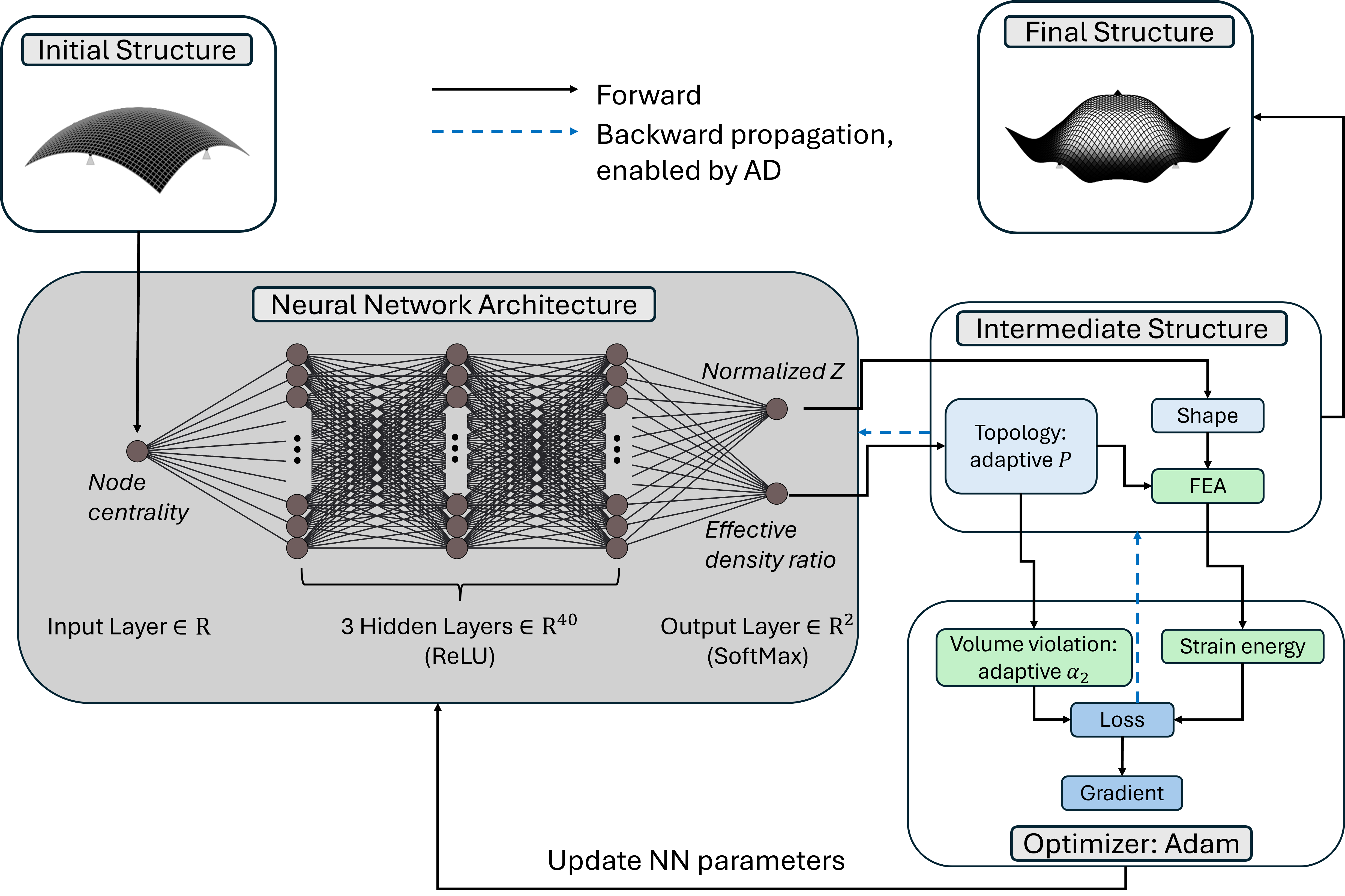}
\caption{NN for structural optimization: flowchart}
\label{fig:nn_flow_2}
\end{figure*}
We leverage the framework presented in Section \ref{sec:NN_FLOW_1} to incorporate NN to conduct simultaneous shape and topology optimization for a shell structure and the flowchart of the approach we take is further shown in Figure \ref{fig:nn_flow_2}. The initial dome-like shell structure is illustrated in Figure \ref{fig:nn_ini}.a and Figure \ref{fig:nn_ini}.b, and it spans 6 m along both X and Y axes with a rise of 1.8 m. It is discretized into 40 by 40 quadrilateral shell elements and there are 1681 nodes in the structural system. The four midpoints of four edges are simply supported, and a point load of 500KN is applied to each of the four corner nodes. All the shell elements have the same sectional properties: thickness is set as 150 millimeters, reference Young’s modulus $E_*$=20000 Gpa, and Poisson’s ratio $\nu$=0.3.\par
A simple NN architecture is used, which is illustrated in Figure \ref{fig:nn_flow_2}. The input layer has one feature, the centrality of a node and we are interested in training a NN to output the optimal shape variable and topology variable for this node. When we feed the trained NN with the centrality values of all the nodes, the NN will output the optimized shape and topology of the structure. The centrality of a node is defined as the summation of its planar distance to all supports on the X-Y plane and the normalized centrality is calculated as follows:
\begin{equation}
    c_i = \frac{\Sigma_{j=i_{b,1}}^{j=i_{b,4}}\sqrt{(X_i-X_j)^2+
    (Y_i-Y_j)^2}}{\max_{i\in(1,2,...,1681)} \Sigma_{j=i_{b,1}}^{j=i_{b,4}}\sqrt{(X_i-X_j)^2+
    (Y_i-Y_j)^2}}
\end{equation}
Where X and Y are the coordinates along the X-axis and Y-axis, respectively; the subscript $i_{b,1}$ to $i_{b,4}$ represents the indices for the supported nodes. The normalized centrality of the shell structure is plotted in Figure \ref{fig:nn_ini}.c where nodes near the structural center have lower centrality values as opposed to the nodes near the corners. The reason to use centrality as the inputs of NN is due to the symmetric plan view of the initial structure: nodes with similar centrality values should have similar density values and position after optimization and NN is able to “learn” this pattern through the minimization of the loss function. \par
The rest of the NN architecture is as follows (Figure \ref{fig:nn_flow_2}). There are three hidden layers and each hidden layer consists of 40 neurons with ReLU (Rectifier Linear Unit) activation functions; the output layer outputs two features of the $i$-th node: the shape variable $p_{S,i}$ which is the normalized Z coordinate as described in Equation \ref{eq:nomarl_z} and the intermediate topology variable which is the effective density ratio to penalize the Young’s modulus as described in Equation \ref{eq:simp_eq}. The upper bound $Z_{max}$ and lower bound $Z_{min}$ for the normalized Z-coordinate is set to be 3 m and 0 m, respectively. We apply the SoftMax activation function to the output layer to ensure the shape variable and topology variable range between zero and one. There are in total 3442 Neural Network parameters $\bm{p}_{NN} \in \mathbb{R}^{3442}$ to be trained whose dimension is similar to the traditional direct optimization approach where there are 1600 topology variables and 1681 shape variables. Here we use Flax as the ML library, which can be replaced by any other ML library one prefers, such as PyTorch or TensorFlow.\par
The NN outputs determine the intermediate shape and topology of the structure and the next step is to conduct FEA based on NN outputs. We first convert the intermediate effective density of nodes to the final topology variable $\bm{p}_T \in \mathbb{R}^{3442}$ for each shell element, which is done by averaging the four nodes’ effective density ratios for each shell element. Standard hat filter with a radius of 1.5m is used to smoothen the final shape variables and effective density values. FEA is then conducted.\par
The loss function $g(\bm{p}_{NN})$ to train the NN is formulated as follows, which consists of a penalized strain energy term and a penalized violation for a material usage constraint, similar to the one used in \cite{Chandrasekhar2021TOuNN:}:
\begin{equation}
    g(\bm{p}_{NN})=\frac{\bm{f}^T\bm{u}}{\alpha_1}+\alpha_2(\frac{\Sigma_{j=1}^{j=1600}p_{T,j}}{V^*}-1)^2
\end{equation}
Where the first term $\frac{\bm{f}^T\bm{u}}{\alpha_1}$ is the penalized strain energy of the system and $\alpha_1$ is taken as the initial strain energy of the system; the second term, $\alpha_2(\frac{\Sigma_{j=1}^{j=1600}p_{T,j}}{V^*}-1)^2$, reflects the violation of the material usage constraints $V^*$ and we set $V^*$ to 800 that approximately constrains the total material volume usage to be 50\% of the initial continuous shell structure. The coefficient $\alpha_2$ is adaptive and gradually increases throughout the training of NN: at first epoch $\alpha_2=0.1$ and gradually increases by 0.05 per epoch. We also set the penalty factor $P$ for shell elements’ Young’s modulus (Equation \ref{eq:simp_eq}) adaptive during the training process: $P$ starts at 2 and increases by 0.06 per epoch until it reaches 8.\par
\begin{figure*}[h]
\centering
\includegraphics[width=1\linewidth]{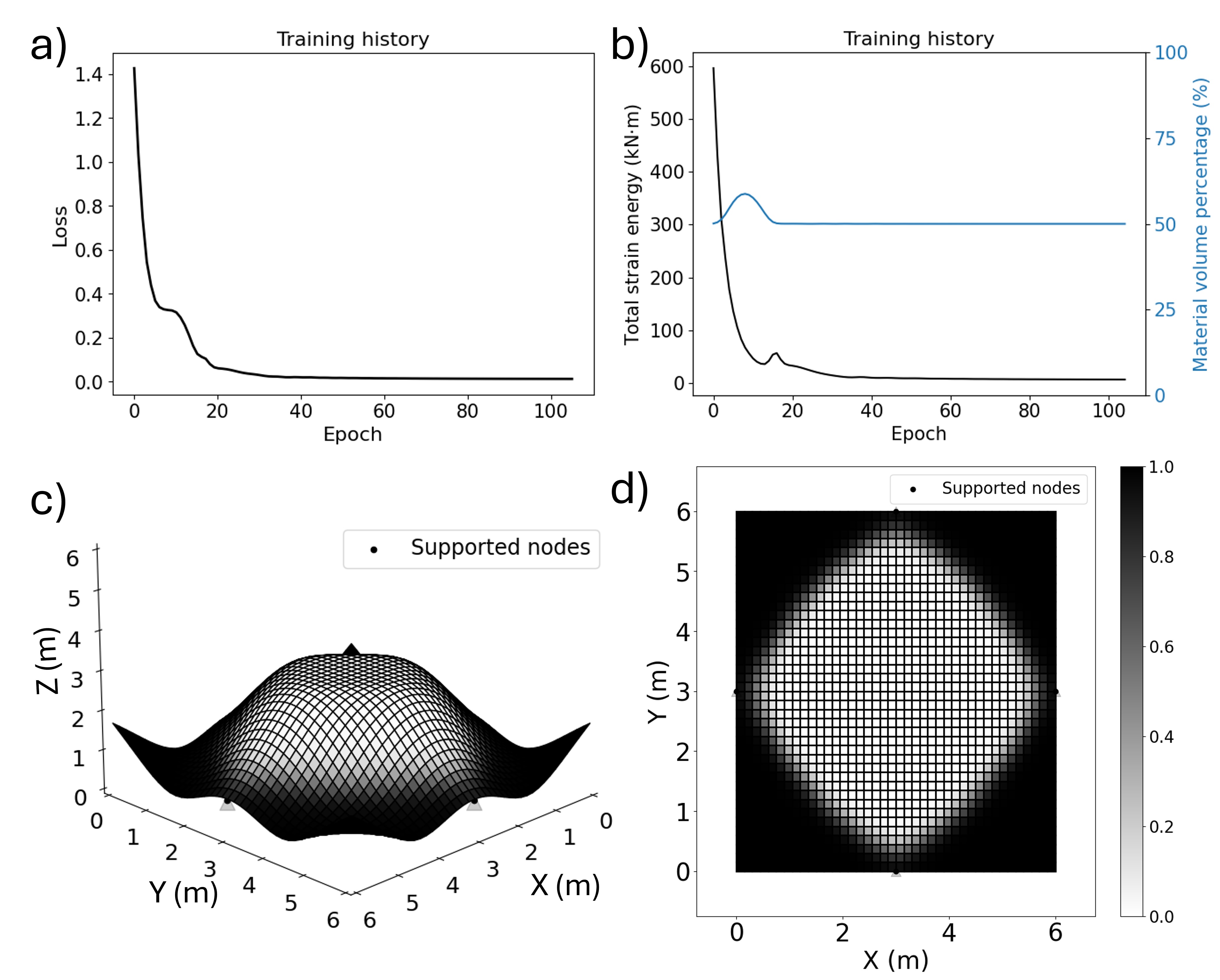}
\caption{NN for structural optimization, results: a) Training history: loss; b) Training history: strain energy and material usage; c) Final structure: perspective view; c) Final structure: plan view with effective density distribution graph where white regions represent void}
\label{fig:nn_final}
\end{figure*}
To train the NN, a robust stochastic gradient descent algorithm, Adam \cite{Kingma2017Adam:}, with a learning rate of 0.01, is used to update the NN parameters $\bm{p}_{NN}$. The maximum epoch is set to be 100. At each epoch, the gradient of the loss function with respect to NN parameters $\bm{p}_{NN}$ is automatically calculated by integrating JAX-SSO and Flax. At the end of the training, the NN outputs the optimized shape and topology of the shell structure.\par
The results are presented in Figure \ref{fig:nn_final}. Figure \ref{fig:nn_final}.a shows the training history of the loss function: the loss function of at the first epoch is 1.43 and it decreases to 0.01 after 100 epochs, which represents a decrease of 99.2\%. Since the loss function is composed of penalized strain energy and volume violation, Figure \ref{fig:nn_final}.b plots their values throughout the training process. It shows that the strain energy is reduced from 595.9 kN$\cdot$m to 7.18 kN$\cdot$m after 100 epochs, showing a reduction of 98.8\%. Meanwhile, the material usage of the structure is successfully constrained under 50\%. The final structure is plotted in Figure \ref{fig:nn_final}.c and Figure \ref{fig:nn_final}.d. The final structure presents a dome-like shape with corner tips cantilevering upwards. In terms of the material usage distribution, the shell elements around the center where there is no load become void. On the contrary, the shell elements near the four corners have the densest material usage and the final material usage pattern (Figure \ref{fig:nn_final}.d) resembles the centrality pattern (Figure \ref{fig:nn_ini}.c). The results show that by using specific inputs to NN, the trained NN is able to optimize the structure in a specific way, even with very simple NN architecture.

\section{Conclusions}\label{sec:con}
This work presents JAX-SSO, a differentiable finite element analysis solver, dedicated to assisting efficient structural optimization. We highlight the features of JAX-SSO as follows:
\begin{itemize}
    \item Automatic sensitivity analysis to assist gradient-based structural optimization, enabled by AD
    \item Support GPU acceleration to boost FEA and sensitivity analysis
    \item Written in Python and JAX, it can be seamlessly integrated with ML libraries for research integrating ML and structural optimization
\end{itemize}
The efficiency of JAX-SSO is illustrated through examples on shape optimization, size optimization, topology optimization and integration with NN. Some limitations of JAX-SSO should be noted, which leads to future direction of further development of JAX-SSO:
\begin{itemize}
    \item The supported element types are limited, more element types should be added to JAX-SSO such as solids
    \item Nonlinearity needs to be addressed in future versions of JAX-SSO
    \item Dynamic analysis module should be considered in the future
\end{itemize}

\section*{CRediT authorship contribution statement}
\textbf{Gaoyuan Wu}: Conceptualization, Methodology, Software, Validation, Formal analysis, Writing – original draft, review \& editing. 
\section*{Declaration of Competing Interest}
The authors declare that they have no known competing financial interests or personal relationships that could have appeared to influence the work reported in this paper.
\section*{Data Availability}
The source code of JAX-SSO is available on its GitHub repository: \par
\href{https://github.com/GaoyuanWu/JaxSSO}{https://github.com/GaoyuanWu/JaxSSO}.




\bibliographystyle{elsarticle-num}
\bibliography{jaxsso}







\end{document}